\documentclass[10pt,a4paper,twocolumn]{IEEEtran}
\normalsize
\usepackage[colorlinks,
linkcolor=black,
anchorcolor=black,
citecolor=black,
urlcolor=black,
]{hyperref}
\usepackage[hyphenbreaks]{breakurl}
\usepackage{diagbox}
\usepackage[english]{babel}
\usepackage{latexsym}
\usepackage{varioref}
\usepackage{mathrsfs}
\usepackage{array}
\usepackage{hhline}
\usepackage{dcolumn}
\usepackage{tabularx}
\usepackage{algorithm}
\usepackage{algorithmic}
\usepackage{calc}
\usepackage{epsfig}
\usepackage[usenames]{color}
\usepackage{amsmath,amssymb,bigdelim}
\usepackage{t1enc}
\usepackage{pstricks,pst-node,psfrag,pst-plot}
\usepackage{graphicx,parskip,amsbsy}
\usepackage{boxedminipage}
\usepackage{indentfirst}
\usepackage{booktabs}
\usepackage[noadjust]{cite}
\usepackage{caption}
\usepackage{multirow}
\usepackage{subfigure}
\usepackage[usenames]{color}
\usepackage{clrscode}
\usepackage{url}
\usepackage{setspace}
\usepackage{pstool}
\usepackage[shortcuts,acronym]{glossaries}
\usepackage[normalem]{ulem}
\newacronym{5G}{5G}{fifth-generation}
\newacronym{MIMO}{MIMO}{multiple-input multiple-output}
\newacronym{SISO}{SISO}{single-input-single-output}
\newacronym{BS}{BS}{base station}
\newacronym{UE}{UE}{user equipment}
\newacronym{PAS}{PAS}{power angular spectrum}
\newacronym{SMPAC}{SMPAC}{sectored MPAC}
\newacronym{SMPAC in abstract}{SMPAC}{sectored multiprobe anechoic chamber}
\newacronym{MPAC}{MPAC}{multiprobe anechoic chamber}
\newacronym{RTS}{RTS}{radiated two stage}
\newacronym{RC}{RC}{reverberation chamber}
\newacronym{HPBW}{HPBW}{half-power-beamwidth}
\newacronym{UCA}{UCA}{uniform circular array}

\newacronym{RF}{RF}{radio frequency}
\newacronym{OTA}{OTA}{over-the-air}
\newacronym{SNR}{SNR}{signal-to-noise-ratio}
\newacronym{mmWave}{mmWave}{millimeter-Wave}
\newacronym{RRM}{RRM}{radio resource management}
\newacronym{PWS}{PWS}{plane wave synthesis}
\newacronym{PFS}{PFS}{prefaded signal synthesis}
\newacronym{CSI}{CSI}{channel state information}
\newacronym{CDL}{CDL}{clustered-delay-line}
\newacronym{DUT}{DUT}{device under test}

\newacronym{VNA}{VNA}{vector network analyzer}
\newacronym{CTF}{CTF}{channel transfer function}
\newacronym{MPC}{MPC}{multipath component}
\newacronym{LoS}{LoS}{line-of-sight}
\newacronym{NLoS}{NLoS}{non line-of-sight}
\newacronym{3GPP}{3GPP}{3rd Generation Partnership Project}

\newcommand{\tabincell}[2]{\begin{tabular}{@{}#1@{}}#2\end{tabular}}
\newcommand{\rem}[1]{}

\newcommand{\addnew}[1]{{\color{black} #1}}
\renewcommand{\baselinestretch}{1}
\providecommand{\texlivekeywords}[1]{\textbf{\textit{Index terms---}} #1}
\begin{document}

\pagestyle{plain}
\title{
Dynamic mmWave Channel Emulation in a Cost-Effective MPAC with Dominant-Cluster Concept} 
\author{ Xuesong Cai,~\IEEEmembership{Member,~IEEE,} Yang Miao, Jinxing Li\\ Fredrik~Tufvesson,~\IEEEmembership{Fellow,~IEEE,} Gert Fr{\o}lund Pedersen, Wei Fan,~\IEEEmembership{Senior Member,~IEEE}  

 \thanks{X. Cai and F.~Tufvesson are with the Department of Electrical and Information Technology, Lund University,
 22100 Lund, Sweden (e-mails: xuesong.cai@eit.lth.se; fredrik.tufvesson@eit.lth.se). The work was started when X. Cai was with the Department of Electronic Systems, Aalborg University, Aalborg 9220, Denmark.

 Y. Miao is with the Radio Systems, Faculty of Electrical Engineering Computer Science and Mathematics (EEMCS), University of Twente, the Netherlands (e-mail: {y.miao@utwente.nl}).

 J. Li is with Huawei Technologies, Shanghai Research Institute, Shanghai 201206, China (e-mail: lijinxing3@huawei.com).


G. F. Pedersen and W. Fan are with the Department of Electronic Systems, Aalborg University, Aalborg 9220, Denmark (e-mails: gfp@es.aau.dk; wfa@es.aau.dk).

}

 %


 %
 %
 %
 %
 %
}

\markboth{IEEE Transactions on Communications}%
{Submitted paper}

\maketitle \thispagestyle{plain}

\begin{abstract}

\Acf{mmWave} massive \acf{MIMO} has been considered as a key enabler for the \acf{5G} communications. It is essential to design and test \acs{mmWave} \acs{5G} devices under various realistic scenarios, since the radio propagation channels pose intrinsic limitations on the performance. This requires emulating a realistic dynamic \acs{mmWave} channels in a reproducible manner in laboratories, which is the goal of this paper. In this contribution, we firstly illustrate the dominant-cluster(s) concept, where the non-dominant clusters in the \acs{mmWave} channels {are pruned}, for \acs{mmWave} \acs{5G} devices applying massive \acs{MIMO} beamforming. This demonstrates the importance and necessity to accurately emulate the \acs{mmWave} channels at a cluster level rather than the composite-channel level. Thus, an \acf{OTA} emulation strategy for dynamic \acs{mmWave} channels is proposed based on the concept of dominant-cluster(s) in a \ac{SMPAC in abstract}. The key design parameters including the probe number and the angular spacing of probes are investigated through comprehensive simulations. A cost-effective switch-circuit is also designed for this purpose and validated in the simulation. Furthermore, a dynamic \acs{mmWave} channel measured in an indoor scenario at 28-30\,GHz is presented, where the proposed emulation strategy is also validated by reproducing the measured reality.

\end{abstract}

\texlivekeywords{Millimeter-wave, massive MIMO, beamforming, dynamic channels and clusters, over-the-air testing, \acf{MPAC}, channel emulation}

\IEEEpeerreviewmaketitle

\section{Introduction}

To meet the needs of immensely increased wireless data consumption driven by a diversity of applications and devices, \ac{5G} networks with even lower latency, higher spectral efficiency, and higher reliability are under deployment. \Ac{mmWave} communication with a large amount of available spectrum is one of the key enablers. As the carrier frequencies increase and antenna apertures decrease significantly, one countermeasure to improve the link margin is massive \ac{MIMO}. A massive \ac{MIMO} device is expected to establish multiple narrow beams to multiple users and dynamically steer the beams as users move or channel conditions change. In such a way, by combining beamforming and spatial multiplexing, a massive \ac{MIMO} system can serve unconventionally large number of terminals in the same time-frequency resources \cite{massiveMIMO,mmWave5G,Sun2014}.
Nevertheless, the \ac{mmWave} radio propagation channel poses many challenges such as severe power losses, scattering and blockage from e.g. human body and moving vehicles, etc. \cite{mmWavePropagation1, mmWavePropagation2, mmWavePropagation3}. The uncontrollable dynamic (time-varying) channels not only limit the coverage, but may also cause serious performance degradation, e.g., due to the loss of user tracking. Therefore, it is important to design and evaluate the \ac{mmWave} devices considering various real-world mmWave channel conditions. 
However, in-field device testing is always expensive and exposed to unpredictable uncertainties. Moreover, \ac{mmWave} massive \ac{MIMO} systems are composed of tens or hundreds of antenna elements as an integrated unit. It is thus neither feasible nor economic to implement conductive testing by connecting each radiating element to \ac{RF} cable at \ac{mmWave} frequencies. Therefore, \ac{OTA} testing which refers to test wireless devices in the laboratory environment without cable-connection has become an essential performance validation procedure for \ac{mmWave} massive \ac{MIMO} devices. In \ac{OTA} testing, the actual dynamic channel condition is mimicked in lab as if the devices were put into use in the real-world. It saves time and money and, most importantly, is reproducible hence provides fair assessment of devices.


The \ac{OTA} testing for \ac{mmWave} massive \ac{MIMO} devices must meet the requirements on signal quality, antenna calibration, demodulation (data throughput performance) and \ac{RRM} \cite{WeiMagazine}. \ac{RRM} refers to the initial access to system, the connection reconfiguration, the handover during mobility, the beam refinement and tracking, and it should be based on realistic \ac{mmWave} channel conditions and dynamic spatial profiles. To meet the above requirements, different \ac{OTA} testing strategies including \ac{RC}, \ac{RTS} method and \ac{MPAC} have been proposed.
In the \ac{RC} approach, metallic stirrers are used in an enclosed metallic cavity to produce a random field variation. As only uniform \ac{PAS} can be emulated in \ac{RC}, little control is available over angular distribution and channel depolarization \cite{RC1, RC2}. Since \ac{mmWave} channels are highly sparse and directive \cite{CHANNELMAS}, i.e., being dominant by a few propagation paths/clusters, \ac{RC} is thus less suitable. Nevertheless, reconfigurable \ac{RC} whose walls support for the reconstruction of controllable 3D \ac{PAS} is under investigation.
For the \ac{RTS} method \cite{6819457,2Stage} which aims to achieve cable connection function without actual \ac{RF} cables, the transfer function between the probe antennas and the antenna ports of \ac{DUT} are effectively calibrated via implementing an inverse matrix in the channel emulator. It is capable of emulating arbitrary dynamic channel. However, the main drawback is that the antenna systems on the \ac{DUT} has to remain static (i.e. non-adaptive to the dynamic channel and therefore beam-locked mode enabled). Therefore, it is not a true end-to-end \ac{OTA} testing method for performance evaluation.
In the \ac{MPAC} approach, the fading emulator synthesizes the fields in the test zone by controlling the excitation and radiation of probes. The \ac{MPAC} setup can emulate arbitrary \ac{PAS} and has easier control over the polarization \cite{MPAC1,MPAC4,MPAC2,MPAC3,MPAC5}.
To address the bi-directional (both uplink and downlink) and 3D spatial channels for testing \ac{mmWave} massive \ac{MIMO} devices, the conventional \ac{MPAC} configuration has been evolved into the 3D \ac{SMPAC} configuration as discussed in \cite{8421660, WeiMagazine}. The sectored configuration refers to one or a few sector(s) of co-located probes deployed on the partial spherical surface with dual-polarized antennas \cite{WeiMagazine}.
The key design parameters include \textit{i)} the measurement range or the distance between \ac{DUT} and probes, \textit{ii)} the number of probes and \textit{iii)} the amount of \ac{mmWave} channel emulator resources. These in turn depend on the major clusters of paths in realistic propagation channels and the desired emulation accuracy. {While the \ac{mmWave} probe antennas can be made cheap (e.g., using patch antennas), the associated \ac{RF} chains connected to the probes including \ac{RF} cables and up/down-frequency converters\footnote{\addnew{Up-frequency converters are required when signals are streamed from the emulator to the \ac{DUT}, and vice versa.}} as well as the baseband components in a channel emulator are very expensive.} Given that the \ac{mmWave} massive \ac{MIMO} channel has a highly sparse and directive angular profile, part of the probes maybe inactive during emulation, a probe selection mechanism with a switch-circuit can be used to reduce cost on \ac{RF} chains while maintaining accuracy. For these advantages, this paper focuses on the \ac{SMPAC}. 

Meanwhile, many measurement campaigns, e.g. \cite{3GPP38901,caidynamic,8713575}, have shown that \ac{mmWave} channels are sparse and mainly power-limited caused by path loss, blockage, etc. Thus beamforming in \ac{mmWave} is essential by transmitting coherent signals thus forming a concentrated field to increase \ac{SNR} or throughput. This in turn, as shown later in Sect.\,\ref{sec:beamformingeffect}, filters the channel. Similar demonstration can be found in \cite{8568555} where the spatial channel at the \ac{UE} side can be significantly simplified if \ac{BS} beamforming operation is applied. In other words, it is not necessary to emulate all the clusters, since some clusters become insignificant with beamforming. Moreover, beams of a \ac{BS} and \acp{UE} should be able to reliably track each other, which means that the dynamic characteristics at a  cluster-level is essential to be emulated.
Although the existing \ac{SMPAC}, e.g. in \cite{WeiMagazine,8421660,8568536}, has several advantages over the conventional \ac{MPAC}, there are still research gaps that need to be addressed considering the above mentioned new \ac{mmWave} features. Limitations include \textit{i)} All the clusters in the \ac{mmWave} channel are emulated. The cost of the associated \ac{RF} resources may be still considerable as a relatively large number of active probes is required if a high emulation accuracy is demanded. \textit{ii)} Since the channel is emulated in a composite manner, cluster-level dynamic characteristics, e.g. of the dominant cluster, may be not well reproduced. However, \ac{mmWave} massive \ac{MIMO} communications mainly rely on this dominant cluster. It is essential to reproduce the dominant cluster/clusters with high emulation accuracy for performance evaluation. \textit{iii)} The switch-circuit with full freedom resultes in high cost and complexity. How to devise a dedicated but low-profiled switching matrix with expected flexibility to support the targeted dynamic channel still needs to be refined.

Overall, the \ac{SMPAC} setup is promising for \ac{OTA} testing of \ac{mmWave} massive \ac{MIMO} devices. However, the performance of such setup on realizing highly dynamic, sparse and directive \ac{mmWave} channels is yet to be enhanced with emphasis on cluster-level behavior and lower cost profile. Whether it is possible to emulate a dynamic selection of a limited number of cluster(s) instead of all clusters, to further reduce the cost on \ac{RF} resources and the complexity of the switch-circuit with dedicated flexibility while maintaining/improving the emulation accuracy and better meeting the testing requirements, is lacking investigation in literature. To fill the above mentioned research gaps, the main contributions of this paper are as follows:
\begin{itemize}

\item {The beamforming effect of \ac{mmWave} massive \ac{MIMO}, i.e. channel simplification, is discussed, which is the basic reasoning for the proposed dominant-cluster(s) concept. This idea has been briefly discussed in the standardization meetings \cite{3GPPdocument1,3GPPdocument2}, though details were not given. Compared to \cite{3GPPdocument1,3GPPdocument2}, we propose two different metrics that correspond to beam management and throughput respectively to evaluate the number of cluster(s) that are dominant for \ac{OTA} emulation.} 

\item {Comprehensive simulations have been performed to find proper design parameters. Especially, to our best knowledge, we for the first time distinguish aligned and nonaligned cluster conditions, which is essential for emulating a dynamic cluster evolving in the angular domain. Moreover, a simpler switching matrix is also proposed with interleaved probe panel design.}

\item {Different from most of the other works solely based on simulations, an indoor dynamic \ac{mmWave} channel measured at 28-30\,GHz is presented. The proposed setup is validated using the realistic measurement data.}




\end{itemize}

The rest of the paper is organized as follows. In Sect.\,\ref{sectII:signalmodel}, the system models of mmWave massive MIMO communication, mmWave propagation channels and channel emulation are elaborated. In Sect.\,\ref{secIII:mpac_simulation}, we discuss the proposed \ac{SMPAC} design for dynamic channel emulation via comprehensive simulations. In Sect.\,\ref{secIV:realistic_channel}, the performance of the design is evaluated by exploiting a realistic indoor dynamic \ac{mmWave} channel. Conclusive remarks are finalized in Sect.\,\ref{section:conclusion}.

\begin{figure*}
 \centering
 \includegraphics[width=0.95\textwidth]{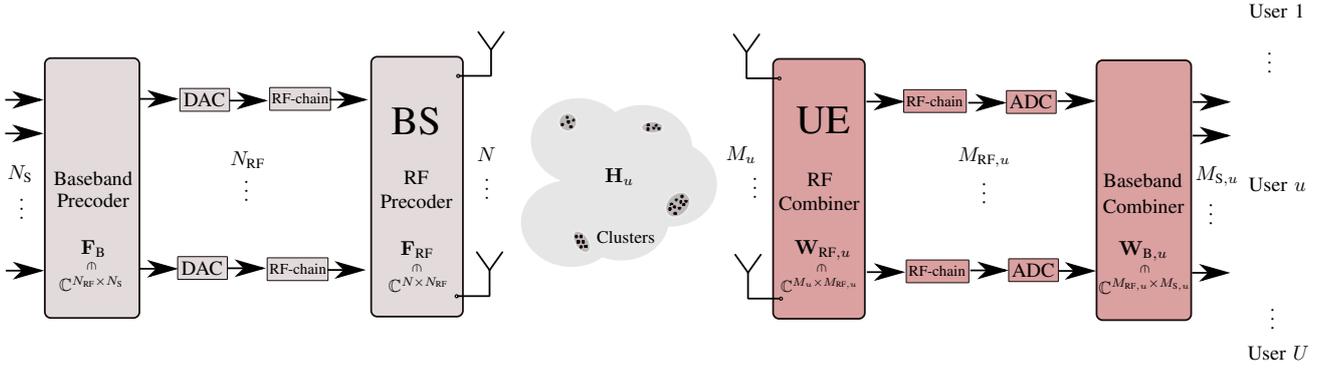}{
 \psfrag{NT}[c][c][0.8]{$N_{\text{}}$}
 \psfrag{NU}[c][c][0.8]{$M_{\text{S},u}$}
 \psfrag{NS}[c][c][0.8]{$N_{\text{S}}$}
 \psfrag{Baseband}[c][c][0.8]{Baseband}
 \psfrag{Pre}[c][c][0.8]{Precoder}
 \psfrag{Combiner}[c][c][0.8]{Combiner}
 \psfrag{RF Chain}[c][c][0.6]{RF-chain}
 \psfrag{2}[c][c][0.7]{$2$}
 \psfrag{N}[c][c][0.8]{$M_{u}$\quad}
 \psfrag{MR}[c][c][0.8]{$M_{\text{RF},u}$}
 \psfrag{MT}[c][c][0.8]{$N_{\text{RF}}$}
 \psfrag{RF}[c][c][0.8]{RF}
 \psfrag{dots}[c][c][0.8]{$\cdots$}
 \psfrag{O}[c][c][0.8]{}
 \psfrag{P}[c][c][0.8]{$\mathbf H_u$}
 \psfrag{Q}[c][c][0.8]{}
 \psfrag{C}[l][l][0.7]{Clusters}
 \psfrag{ADC}[c][c][0.8]{ADC}
 \psfrag{DAC}[c][c][0.8]{DAC}
 \psfrag{Rx}[c][c][1.6]{UE}
 \psfrag{Tx}[c][c][1.6]{BS}
 \psfrag{User 1}[c][c][0.8]{User 1}
 \psfrag{User k}[c][c][0.8]{User $u$}
 \psfrag{User K}[c][c][0.8]{User $U$}
 \psfrag{FBU}[c][c][0.8]{$\mathbf F_\text{B} $}
 \psfrag{FRU}[c][c][0.8]{$\mathbf F_{\text{RF}} $}
 \psfrag{WBU}[c][c][0.8]{$\mathbf W_{\text{B},u}$}
 \psfrag{WRU}[c][c][0.8]{$\mathbf W_{\text{RF},u}$}
 \psfrag{FBD}[c][c][0.7]{$\mathbb{C}^{N_{\text{RF}}\times  N_\text{S}}$}
 \psfrag{FRD}[c][c][0.7]{$\mathbb{C}^{N_\text{}\times  N_\text{RF}}$}
 \psfrag{WRD}[c][c][0.65]{$\mathbb{C}^{ M_{u}\times M_{\text{RF},u}}$}
 \psfrag{WBD}[c][c][0.7]{$ \mathbb{C}^{ M_{\text{RF},u}\times M_{\text{S},u}}$}
 \psfrag{in}[c][c][0.6][-90]{$\in$}
 }
 \caption{System model of the wireless communication system.\label{fig:radio_channel}}
\end{figure*}

\section{System Model\label{sectII:signalmodel}}

In this section, we introduce the model of the \ac{mmWave} massive \ac{MIMO} communication system, which is useful to help readers understand the theory and practice as well as the resulting limitations upon \ac{OTA} channel emulation. The signal models of the \ac{MIMO} channel and \ac{OTA} channel emulation are also elaborated on. Finally, the effect of beamforming in massive \ac{MIMO} is illustrated for the dominant-clusters concept.



\subsection{Wireless Communications Model\label{sec:precoder}}

\addnew{In this subsection, we provide a short review for the massive \ac{MIMO} communication using hybrid array configuration \cite{7448873,8241348,6979963}.} 
As illustrated in Fig.\,\ref{fig:radio_channel}, we exemplify the concept with a single-cell downlink wireless communication scenario where one \ac{BS} and $U$ \acp{UE} are considered. The \ac{BS} has $N_{\text{}}$ antenna elements and $N_{\text{RF}}$ RF-chains, and $N_{\text{S}}$ data streams are transmitted to serve $U$ users ($N_{\text{S}}\leq N_{\text{RF}}\leq N_{\text{}}$). The $u$th user has $M_{u}$ antenna elements and $M_{\text{RF},u}$ \ac{RF}-chains, respectively, and $M_{\text{S},u}$ data streams are intended for this user ($M_{\text{S},u}\leq M_{\text{RF},u}\leq M_{u}$). At the \ac{BS} side, data symbols $\mathbf S(f,t) \in \mathbb{C}^{N_{\text{S}}\times  1}$ to be transmitted at the subcarrier $f$ and time instant $t$ are firstly precoded with the baseband digital precoding matrix $\mathbf F_\text{B} \in \mathbb{C}^{N_{\text{RF}}\times  N_\text{S}}$ and then precoded by the \ac{RF} precoding matrix $\mathbf F_\text{RF} \in \mathbb{C}^{N_{\text{}}\times  N_\text{RF}}$ in the \ac{RF} domain. That is, the complex-equivalent symbol-vector $\mathbf X(f,t) \in \mathbb{C}^{N_{\text{}}\times  1}$ sent is formatted as
\begin{align}
 \mathbf X(f,t) =\mathbf F_\text{RF}(f,t) \mathbf F_\text{B}(f,t) \mathbf S(f,t)\label{eq:precoding}.
\end{align}
At the \ac{UE} side, the received signal $\mathbf Y_u(f,t) \in \mathbb{C}^{M_{u}\times  1}$ at the $M_{u}$ antenna ports of the $u$th user is
\begin{align}
 \mathbf Y_u(f,t) = \mathbf H_{u}(f,t) \mathbf X(f,t) + \mathbf N(f,t) \label{eq:channel}
\end{align} where $\mathbf N(f,t)$ represents white Gaussian noise, and $\mathbf H_u(f,t) \in \mathbb{C}^{M_{u}\times N_\text{}}$ is the \ac{MIMO} channel transfer matrix for the $u$th user, which is defined later in (\ref{eq:geometrical_model}). Similarly, $\mathbf Y_u(f,t)$ is firstly combined in the \ac{RF} domain by applying the \ac{RF} combining matrix
$\mathbf W_{\text{RF},u} \in \mathbb{C}^{M_{u}\times  M_{\text{B},u}}$ and then combined in the baseband domain with the digital combining matrix $\mathbf W_{\text{B},u} \in \mathbb{C}^{M_{\text{RF},u}\times  M_{\text{S},u}}$. Therefore, the received data symbol vector $\mathbf Z_u(f,t)$ at subcarrier $f$ after signal-processing is
\begin{align}
 \mathbf Z_u =  \mathbf W_{\text{B},u}^{H}  \mathbf W_{\text{RF},u}^{H} \mathbf H_u \mathbf F_\text{RF}  \mathbf F_\text{B} \mathbf S  + \mathbf W_{\text{B},u}^{H}  \mathbf W_{\text{RF},u}^{H} \mathbf N  \label{eq:combiner}
\end{align} where $(f,t)$ is omitted for notation conciseness.
\addnew{It is known from \eqref{eq:combiner} that the adaptive $\mathbf W_{\text{RF},u}$ and $\mathbf F_\text{RF}$ make it infeasible to reproduce the MIMO channel matrix $\mathbf W_{\text{RF},u}^{H} \mathbf H_u \mathbf F_\text{RF}$, e.g. using the \ac{RTS} method. The proper solution for hybrid massive MIMO should be reproducing $\mathbf H_u$.} \addnew{Moreover, to reduce the system overhead in estimating and/or feedbacking the instataneous channel state information, the major approach in practice to enable \ac{mmWave} massive \ac{MIMO} is that the \ac{BS} and \ac{UE} perform joint beam-sweeping using predefined codebooks and then choose the best beam for communication, e.g. as specified in the standardization document \cite{3GPPNR}, which means that the dominant cluster/clusters in the channel is/are essential.}

\subsection{MIMO Channel Model}
The channel transfer matrix $\mathbf H(f,t)$ in (\ref{eq:channel}) is attributed to both the radio propagation environment and the antenna radiation patterns of \ac{BS} and \ac{UE}.\footnote{Without loss of generality, $u$ is omitted for the sake of conciseness.} It is also time-dependent for dynamic channels {as the propagation delays, complex polarimetric gains, angles, Doppler frequencies, etc. of multipath components are evolving.} The widely applied geometrical channel model for $\mathbf H(f,t)$ can be formatted as
\begin{equation}
 \begin{aligned}
  \mathbf H(f,t) = \sum_{\ell=1}^{L(t)} \mathbf G_\text{Rx}(f, -\mathbf k_{\ell}^{\text{Rx}}(t))  \mathbf A_\ell(f,t)  \mathbf G^{T}_\text{Tx}(f, \mathbf k_{\ell}^{\text{Tx}}(t)) \\ \times \exp\{j2\pi \int^{t} \nu_\ell(t^\prime) dt^\prime\}\exp\{-j2\pi f \tau_\ell(t)\}
  \label{eq:geometrical_model}
 \end{aligned}
\end{equation} with $\mathbf A_\ell(f,t) \in \mathbb{C}^{2\times 2} $ as the polarimetric amplitude matrix
\begin{equation}
\begin{aligned}
  \mathbf A_\ell(f,t) = \left[ {\begin{array}{cc}
      \alpha_\ell^{aa}(f,t) & \alpha_\ell^{ab}(f,t) \\
      \alpha_\ell^{ba}(f,t) & \alpha_\ell^{bb}(f,t) \\
     \end{array} } \right]
     \label{eq:polmatrix}
\end{aligned}
\end{equation}
where $L$ is the total path number, $\tau_\ell$ and $\nu_\ell$ indicate the propagation delay and Doppler frequency for the $\ell$th path, $a$ and $b$ represent the polarization pair of the transmit and receive antennas, and $\alpha_\ell^{\diamond\star}$ are the complex amplitudes for transmitted polarization $\diamond$ and received polarization $\star$. Furthermore, $\mathbf k_{\ell}^{\text{Tx/Rx}}$ is the wave vector in the departure/arrival direction at the Tx/Rx side for the $\ell$th path, and
$\mathbf G_\text{Tx}\in \mathbb{C}^{N_{\text{}}\times  2}$  and $\mathbf G_\text{Rx}\in \mathbb{C}^{M_{\text{}}\times  2}$ represent the polarimetric antenna pattern matrices for Tx and Rx defined to a common phase center, respectively. Moreover, the first column and the second column in $\mathbf G_\text{Tx}$ (and $\mathbf G_\text{Rx}$) are antenna pattern vectors of $a$ and $b$ polarizations, respectively. With uplink transmission considered, we have
$\mathbf G_\text{Tx}\in \mathbb{C}^{M_{\text{}}\times  2}$  and $\mathbf G_\text{Rx}\in \mathbb{C}^{N_{\text{}}\times  2}$. Note that (\ref{eq:geometrical_model}) is based on the assumption of plane-wave propagations. When considering spherical wavefronts \cite{8713575,7981398,xuesong_tap}, the model gets more complicated as the polarimetric complex gain of one path varies among antenna elements, and the Doppler frequency also change with respect to different antenna pairs. The spherical wavefront is out of the scope of this paper. Readers are referred to \cite{8713575,7981398,xuesong_tap,Metis} for the corresponding spherical-propagation models.


The purpose of channel emulation is to reproduce the \ac{MIMO} channel, i.e. the channel transfer matrix $\mathbf H$, in a controllable and repeatable way in laboratory to test devices. In traditional conducted \ac{MIMO} emulation, the Tx antenna ports and Rx antenna ports are connected to the input ports and output ports of the channel emulator, respectively. The channel $\mathbf H$ is generated in the fading emulator and multiplied with the input $\mathbf X$, and the resulting signal $\mathbf Y$ is fed to the Rx device \cite{8421660}. However, as explained in the introduction, conventional conducted testing is no longer applicable for \ac{mmWave} devices due to lack of antenna connectors. Highly integrated RF circuits and antenna designs are inevitable at \ac{mmWave} bands due to concern of cost, size and loss. Alternatively, \ac{OTA} testing by exploiting \ac{SMPAC} method has been considered as the most appropriate strategy for the performance evaluation of \ac{mmWave} massive \ac{MIMO} devices.

\subsection{\ac{SMPAC} \ac{OTA} Emulation}
\begin{figure}
  \centering
  \includegraphics[width=0.45\textwidth]{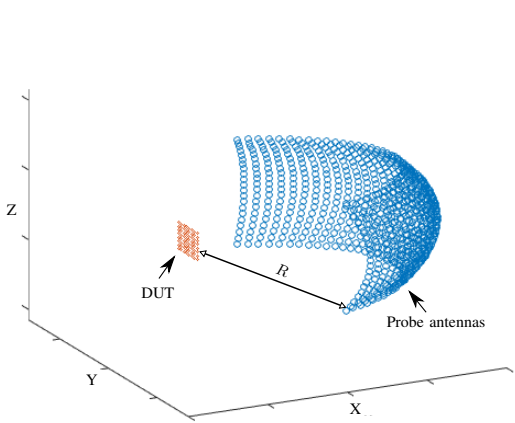}
  {
  \psfrag{x}[c][c][0.6]{X}
  \psfrag{Y}[c][c][0.6]{Y}
  \psfrag{Z}[c][c][0.6]{Z}
  \psfrag{DUT}[c][c][0.6]{DUT}
  \psfrag{D}[c][c][0.6]{Probe antennas}
  \psfrag{R}[c][c][0.6]{$R$}
  }
  \caption{Sectored probe configuration.}
  \label{fig:probes}
\end{figure}

\begin{figure}
  \centering
  \includegraphics[width=0.45\textwidth]{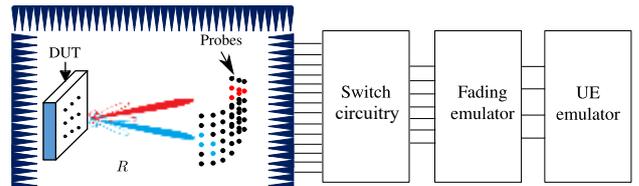}{\psfrag{A}[c][c][0.7]{ \begin{tabular}{@{}c@{}} Switch\\circuitry\end{tabular} }
  \psfrag{B}[c][c][0.7]{ \begin{tabular}{@{}c@{}} Fading\\emulator\end{tabular} }
  \psfrag{C}[c][c][0.7]{ \begin{tabular}{@{}c@{}} \ac{UE}\\emulator\end{tabular} }
  \psfrag{DuT}[c][c][0.6]{DUT}
  \psfrag{D}[c][c][0.6]{Probes}
  \psfrag{R}[c][c][0.6]{$R$}
  }
  \caption{A \ac{SMPAC} system setup.}
  \label{fig:smpac}
\end{figure}

The enhanced beamforming capacity in 3D (both azimuth and elevation) is advantageous to overcome the high path loss in \ac{mmWave} bands and mitigate the interference among users. This in turn requires the \ac{OTA} emulation to support 3D propagation, thus the 2D probe configuration \cite{MPAC1,ictpekka} applied for LTE 4G communications is no longer adequate for \ac{mmWave} Massive \ac{MIMO} devices. A \ac{SMPAC} setup was thus first proposed in \cite{8421660} as illustrated in Fig.\,\ref{fig:probes}. A large number of probes with approximately the same distance $R$ to the \ac{DUT} at the coordinate center and certain angle spacings among them are used to cover a sector of the whole sphere. This is based on the reasonable assumption that the propagation paths of \ac{mmWave} propagations are confined in a certain angle-range/sector for most scenarios, and the hardware cost can be decreased with less probes used. In other words, the \ac{SMPAC} setup is a compromise between the requirement of 3D emulation and the hardware cost. As illustrated in Fig.\,\ref{fig:smpac}, we exemplify the underlying mechanism for the \ac{SMPAC} emulation by the uplink transmission where the \ac{DUT} (Rx) is \ac{BS} and Tx is the \ac{UE} (or \ac{UE} emulator). 
 The setup contains an anechoic chamber, a number of probes with $K$ of them active, a fading emulator, a \ac{UE} emulator and a switch matrix to connect the $K$ output ports of the fading emulator to the desired $K$ active probes. In this setup, the channel transfer function $\mathbf H$ is mainly reproduced by exploiting the fading emulator and probe configuration. Specifically, the channel transfer function $\mathbf H_{\text{re.}}$ reproduced by the setup is formatted as
\begin{equation}
\begin{aligned}
  \mathbf H_{\text{re.}}(f,t) = \mathbf C(f,t) \mathbf E(f,t)
     \label{eq:emulator_H}
\end{aligned}
\end{equation} where $\mathbf C \in \mathbb{C}^{N \times K}$ is the channel transfer matrix due to the chamber from $K$ probes to the $N$ \ac{DUT} antennas, and $\mathbf E \in \mathbb{C}^{K \times M}$ is the channel transfer matrix attributed to the fading emulator between the $M$ ($M_{u}$) \ac{UE} antennas and the $K$ probes.

With the geometrical distribution of the $K$ probes and $N$ \ac{DUT} antennas known, the $(n,k)$th element $C_{n,k}$ of $\mathbf C$ is determined as
\begin{equation}
\begin{aligned}
   C_{n,k}(f) = \mathbf G_{\text{Rx},n}(f, - \mathbf k_{n,k})  \mathbf G_k^{T}(f, \mathbf k_{n,k})
  \\\times\sqrt{PL(f,d_{n,k})} \exp\{-j ||\mathbf k_{n,k}|| d_{n,k}\}
  \label{eq:chamber_channel}
\end{aligned}
\end{equation} where $\mathbf G_{\text{Rx},n}\in \mathbb{C}^{1 \times 2}$ and $\mathbf G_k \in \mathbb{C}^{1 \times 2}$  are the polarimetric antenna radiation patterns for the $n$th \ac{DUT} antenna and the $k$th probe, respectively, and $\mathbf k_{n,k}$, $d_{n,k}$ and $PL(f,d_{n,k})$ are the wave vector, distance and path loss of the propagation between the $k$th probe and the $n$th \ac{DUT} antenna, respectively. It is worth noting that due to the propagation between the
$k$th probe and the $n$th \ac{DUT} antenna is line-of-sight, the polarization matrix is an identity matrix thus omitted in \eqref{eq:chamber_channel}, {and that the time dependence of $\mathbf C$ can be obtained from the distribution variation of selected $K$ probes.} By prober selecting the probe locations, the required \ac{PAS} of the channel can be generated.

Since the temporal and Doppler characteristics of the channel cannot be embedded in $\mathbf C$, they are created by the fading emulator. The channel transfer matrix $\mathbf E$ executed in the fading emulator has the property that its $k$th row $\mathbf E_k$ can be formatted as
\begin{equation}
 \begin{aligned}
  \mathbf E_k(f,t) = \sum_{\ell=1}^{L}  w_{\ell,k} \mathbf G_{\text{FE},k} \mathbf A_{\ell,k} \mathbf G^{T}_\text{Tx}(f, \mathbf k_{\ell}^{\text{Tx}}) \\ \times \exp\{j2\pi \nu_\ell t\}\exp\{-j2\pi f \tau_\ell\}
  \label{eq:fading_emulator_channel}
 \end{aligned}
\end{equation} where $\mathbf G_{\text{FE},k}\in \mathbb{C}^{1 \times 2}$ is the polarimetric pattern configured for the $k$th probe in the fading emulator, and $\mathbf A_{\ell,k}$ and $w_{\ell,k}$ are the polarization matrix and weight of the $\ell$th path for the $k$th probe. By examining \eqref{eq:fading_emulator_channel}, it can be known that the temporal and Doppler characteristics of all the $L$ paths are passed to all the $K$ probes by including the term $\exp\{j2\pi \nu_\ell t\}\exp\{-j2\pi f \tau_\ell\}$. Meanwhile, the polarization characteristics are guranteed by properly setting $\mathbf G_{\text{FE},k}$ and $\mathbf A_{\ell,k}$ in \eqref{eq:fading_emulator_channel}, which also depends on the $\mathbf G_k$ in \eqref{eq:chamber_channel}. More specifically, the polarization matrix of the $\ell$th path between \ac{UE} and \ac{BS} (\ac{DUT}) passed by the $k$th probe can be obtained as
\begin{equation}
 \begin{aligned}
   \mathbf A_{\text{re.},\ell,k} = \mathbf G_k^{T} \mathbf G_{\text{FE},k} \mathbf A_{\ell,k}.
  \label{eq:polarmatirx_generated}
 \end{aligned}
\end{equation} Since both the ranks of $\mathbf G_k$ and $\mathbf G_{\text{FE},k}$ are $1$ and the rank of $\mathbf A_\ell$ in \eqref{eq:geometrical_model} is usually $2$, at least two probes are required to reproduce $\mathbf A_\ell$ as
$(\mathbf A_{\text{re.},\ell,k_1} + \mathbf A_{\text{re.},\ell,k_2})$. {One direct example is to co-locate two linearly-polarized probes that serve two polarizations, respectively. In this case, both $\mathbf G_{k_1}$ and $\mathbf G_{\text{FE},k_1}$ can be $[1,0]$, while both $\mathbf G_{k_2}$ and
$\mathbf G_{\text{FE},k_2}$ can be $[0,1]$, and $\mathbf A_{\ell,k_1/k_2}$ can be $\mathbf A_\ell$.} Moreover, as different paths have different angle of arrivals, the weights $\mathbf W_\ell = \{w_{l,k};k=1,\cdots,K\}$ are optimized together with the locations of the $K$ probes to control the spatial characteristics.

Ideally, to exactly reproduce $\mathbf H$, one can execute $\mathbf E$ in the fading emulator as
\begin{align}
  \mathbf E = \mathbf C^{-1} \mathbf H.
\end{align} However, the difficulty lies in the fact that $\mathbf C$ is typically non-measurable, e.g., due to the hybrid structure.
Moreover, what matters is the statistical behaviour of the geometry-based stochastic channel rather than its instantaneous snapshots \cite{8421660}. Alternatively, the purpose is to reproduce a statistically similar $\mathbf H_{\text{re.}}$ to $\mathbf H$. As discussed above, the statistical behaviour in the temporal, Doppler and polarization domains can be mostly executed in the fading emulator, the challenge is to control the statistical spatial behaviour in the anechoic chamber. This is achieved by properly choosing the $K$ probes and setting $\mathbf W_\ell$. \Ac{PFS} \cite{MPAC1,MPAC4} and \ac{PWS} \cite{MPAC2,MPAC3,MPAC5} approaches can be applied, between which \ac{PFS} is preferable and exploited in this work as the \ac{PWS} requires strict phase coherence to emulate plane-waves in the test-zone.


\subsection{Beamforming Effect on the Channel Emulation\label{sec:beamformingeffect}}


In massive \ac{MIMO} communications, both beamforming and spatial multiplexing can be exploited. 
We distinguish the two terms the same way as in \cite{Sun2014}. That is, beamforming indicates a classically steered beam, while spatial multiplexing indicates that an outgoing signal stream is divided into independent substreams and sent in parallel through the same radio channel. It can, for both theoretical and practical reasons, be argued that beamforming will be applied predominantly in \ac{mmWave} massive \ac{MIMO} communications. \textit{i)} Numerous channel measurement campaigns have shown that the high attenuation at \ac{mmWave} bands mostly results in  power-limited channels. Beamforming is thus necessary to provide sufficient link budget \cite{Sun2014,Giordani2019}, and larger capacity could be achieved with higher order modulations due to increased \ac{SNR} \cite{Sun2014}. \textit{ii)} It is usually required for spatial multiplexing that the \ac{CSI} is available at the Tx side. However, the hybrid structure and large number of antennas pose significant challenges and overheads in \ac{CSI} acquisition \cite{8241348,Sun2014}. Alternatively, beam management with beamforming aiming to select the best beam pair between a \ac{BS} and a \ac{UE} requires less and easier feedback, e.g. as proposed in the \ac{3GPP} standard
\cite{Giordani2019,3GPPNR}. \textit{iii)} Due to hardware limitations, it can be expected that the analog structure is preferable for \ac{UE}s in the early evolution of \ac{mmWave} communication systems. This further hinders the application of spatial multiplexing for a single user.

\begin{figure}
  \centering
  \includegraphics[width=0.4\textwidth]{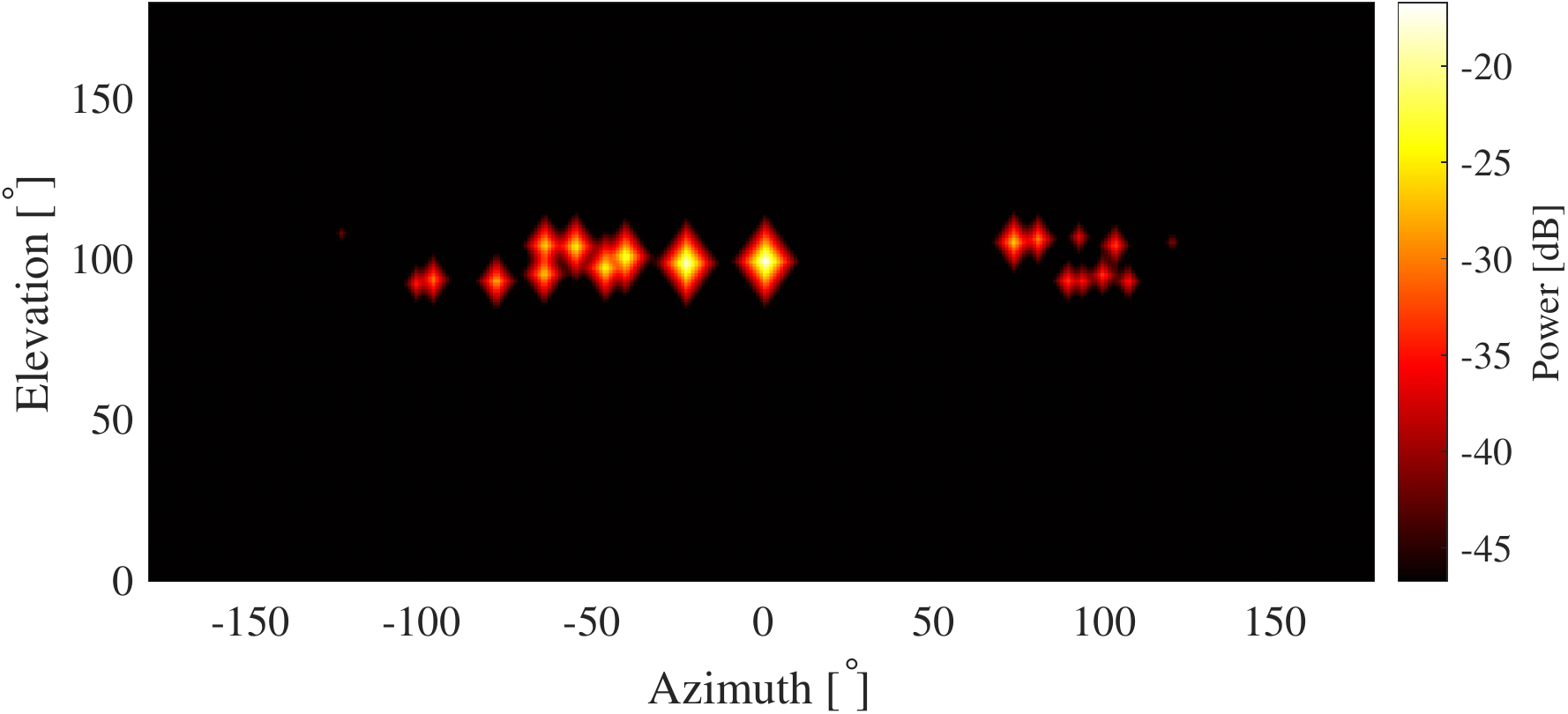}
  \caption{The \ac{PAS} of {CDL} C.}
  \label{fig:pas_cdl_c}
\end{figure}

\begin{figure}
  \centering
  \subfigure[]{\includegraphics[width=0.38\textwidth]{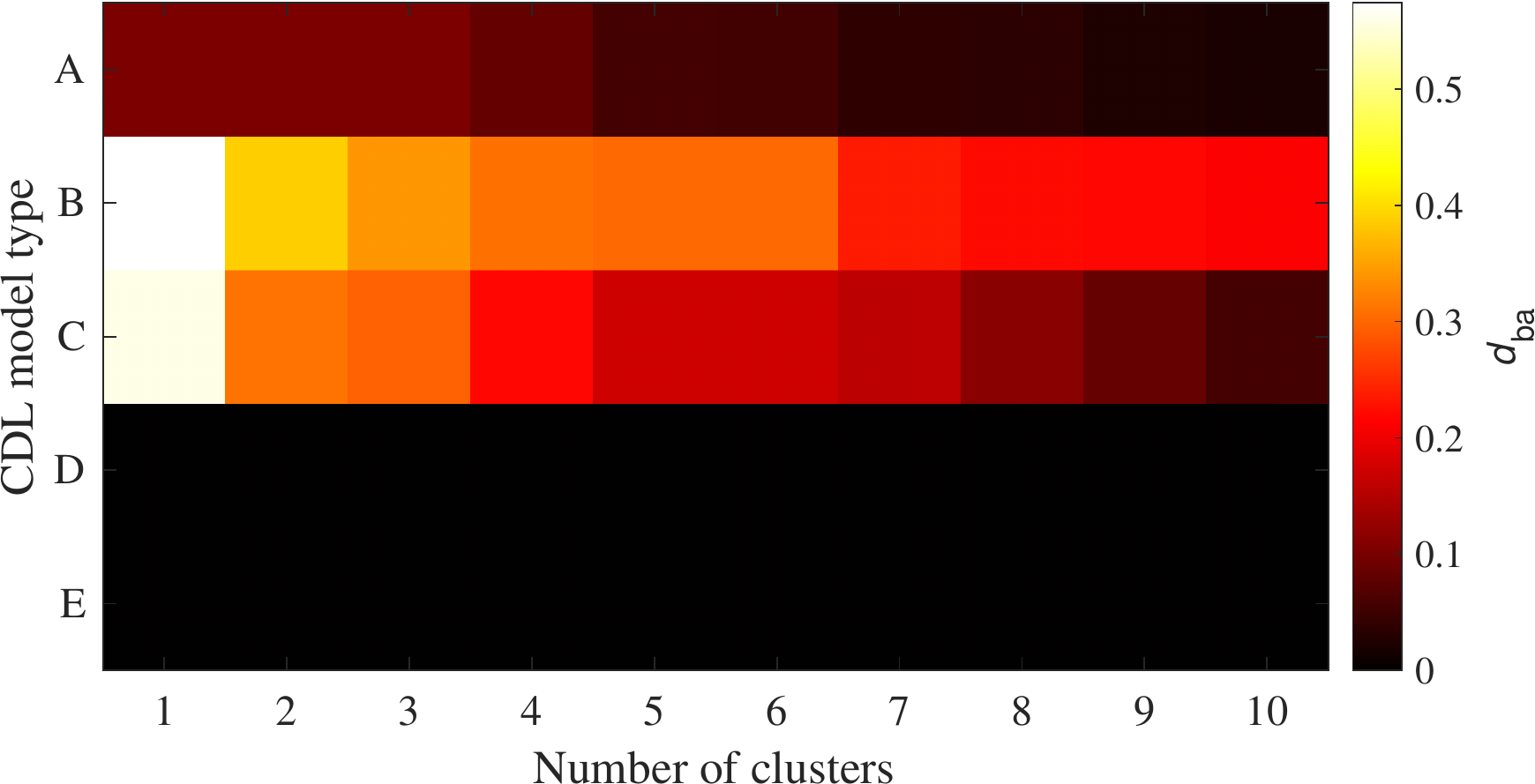}}
  \subfigure[]{\includegraphics[width=0.38\textwidth]{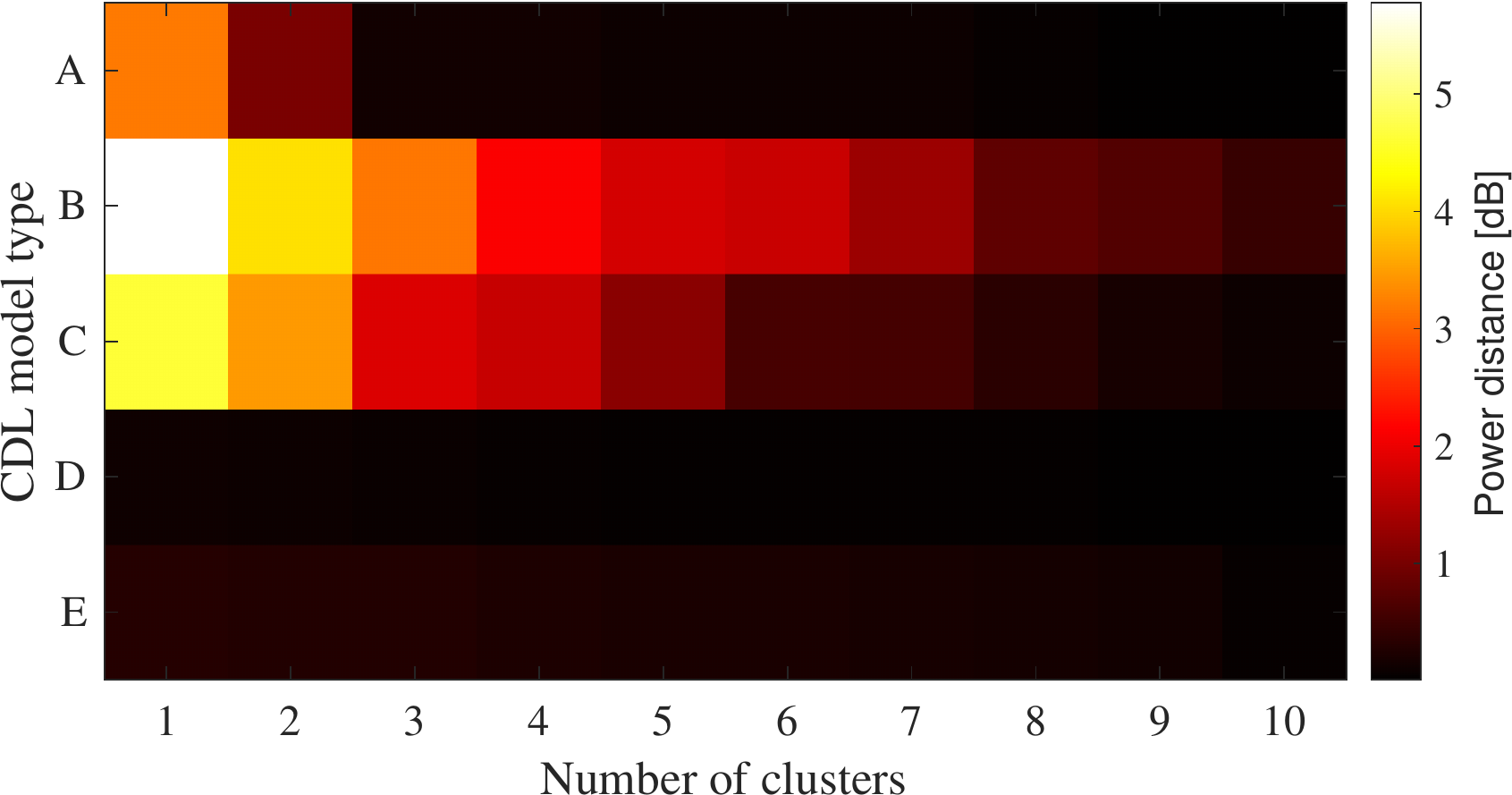}}
  \caption{The evaluation of the number of dominant clusters for an $8\times8$ \ac{DUT} array with beamforming. (a) Total variation distance of beam allocation distributions. (b) Mean power difference of directed beams. }
  \label{fig:channel_prune}
\end{figure}



{It is clear that in the beamforming mode, the propagation channel will be spatially filtered. That is, although multiple clusters may exist in the channel, the directed and narrow beam makes clusters in the other directions insignificant to the communication link. Nevertheless, it is worth noting that clusters fade independently. If the DUT is in beam tracking mode, i.e., following several clusters and always tracking the strongest one, the beam may not be always directed to the cluster with a maximum-power \ac{PAS}. Thus, there may be one or several clusters that can be considered dominant, and they should not be pruned in \ac{OTA} emulation. To illustrate the effect of beamforming on \ac{mmWave} channels, five different representative \ac{3GPP} \ac{CDL} models (A-E) specified in \cite{3GPP38901} are exploited, where models D-E and A-C correspond to \ac{LoS} and \ac{NLoS} scenarios, respectively. As an example, the \ac{PAS} of CDL model~C is illustrated in Fig.\,\ref{fig:pas_cdl_c}. Since it is important to evaluate the beam management performance and the throughput of the \ac{DUT}, the beam allocation distribution and the mean power of all directed-beams\footnote{Directed beam means the beam with the maximum power during beam sweeping.} are used respectively to compare the original channel and the pruned channel with different number of dominant clusters included. Specifically, the \textit{total variation distance of beam allocation distributions} [see Sect.\,\ref{sec:metrics}(3)] and the distance of the mean powers are calculated for each of the five scenarios considering an 8$\times$8 \ac{DUT} array.
 As illustrated in Fig.\,\ref{fig:channel_prune}, it can be observed that the LoS path only is always enough for the LoS channel scenarios D-E. However, for the NLoS channel scenarios, several dominant clusters are needed to achieve a satisfactory emulation performance. For example for CDL~A, one or two dominant clusters are quite fine. For CDL~C, five clusters may be needed. Since the channel can be pruned by considering only the dominant cluster(s), it is important to reproduce each cluster as accurately as possible especially when they are evolving in the angular domain in dynamic scenarios. In the sequel, investigations on how to accurately emulate a dynamic cluster are elaborated. Multiple clusters can be emulated together by emulating individual dynamic clusters accurately. Compared to the method of emulating the composite channel as a whole, the proposed method emphasizes the emulation accuracy of a few dominant clusters (if not one), which is important for mm-wave massive MIMO communications and also saves expensive resources. Moreover, the dynamic behavior of a cluster can be well considered.}

\section{SMPAC Design for \ac{OTA} Channel Emulation\label{secIII:mpac_simulation}}

In this section, probe allocation and weighting in \ac{SMPAC} setup are discussed first. Four metrics (including total variation distance of \ac{PAS}, spatial correlation error, beam peak distance and total variation distance of beam allocation distributions) are then used in the simulation to evaluate how the \ac{SMPAC} design parameters, e.g., probe number and angle spacing, affect the emulation performance for a single-cluster channel. Moreover, a novel cost-effective design for probes and switching strategy is discussed to emulate the dynamic channels.

\subsection{Probe Allocation and Weighting}
In \ac{PFS}, independent and identically distributed (i.i.d.) fading sequences (e.g. Rayleigh distribution \cite{MPAC1}) are generated at the selected $K$ probes. By applying probe weights $w_k$ for each of the $K$ probes, it is expected to reproduce the target \ac{PAS} $P(\Omega)$.\footnote{Note that the subscript $\ell$ is omitted in $w_{\ell,k}$ because the weights for different paths in the same cluster should be the same.} The weights $w_k, k=1,\cdots,K$ are obtained by minimizing the square error between the target spatial correlations and reproduced spatial correlations. Specifically, with a target \ac{PAS} $P(\Omega)$, the spatial correlation between any two (the $a$th and $b$th) \ac{DUT} antennas can be formatted as
\begin{equation}
 \begin{aligned}
  \rho_{a,b} = \frac{\oint P(\Omega) \exp\{j \mathbf k_\Omega \cdot (\mathbf a - \mathbf b)\}  d \Omega}{\oint P(\Omega) d\Omega}
  \label{eq:target_spatial_corr}
 \end{aligned}
\end{equation} where $\mathbf a$ and $\mathbf b$ are the location vectors for the $a$th and $b$th \ac{DUT} antennas, respectively, and $\mathbf k_\Omega$ is the wave vector in the direction of space angle $\Omega$. With $K$ probes selected, the reproduced spatial correlation between the $a$th and $b$th \ac{DUT} antennas can be calculated as \cite{8421660,MPAC1}
\begin{equation}
 \begin{aligned}
  \hat{\rho}_{a,b} = \frac{\sum_{k=1}^{K} w_k^2 PL(d_{a,k}) PL(d_{b,k}) \exp\{j||\mathbf k|| (d_{a,k} - d_{b,k})\}}{\sqrt{\sum_{k=1}^{K}  w_k^2 PL^2(d_{a,k}) \sum_{k=1}^{K} PL^2(d_{b,k})}}.
  \label{eq:realized_spatial_corr}
 \end{aligned}
\end{equation} The numerator in \eqref{eq:realized_spatial_corr} can give some insight that the cross terms have to vanish with i.i.d. fading sequences applied for different probes, and the denominator in \eqref{eq:realized_spatial_corr} is simply a normalization factor. The optimized $\mathbf W$ is then obtained by minimizing the square error between the target correlation function and the reproduced correlation function as
\begin{equation}
 \begin{aligned}
  \mathbf W = \arg \min_{\mathbf W} \sum_{a=1}^{N} \sum_{b=1}^{N} | \rho_{a,b} -  \hat{\rho}_{a,b}|^2.
  \label{eq:determining_weights}
 \end{aligned}
\end{equation} It is worth noting that the probe weighting and allocation are actually a joint optimization problem. In this work, we attempt to select the best $K$ probes within two steps. First, a relatively large probe area is considered active for the \ac{SMPAC} emulation, and their weights are calculated according to \eqref{eq:determining_weights}. Then the probes with the $K$ highest $|w_k|^2$ are selected to be active only, and the weights for the $K$ probes are optimized again using \eqref{eq:determining_weights}.

\subsection{Performance Evaluation Metrics\label{sec:metrics}}
The critical \ac{SMPAC} design parameters include the range $R$ between DUT and probes, the angular spacing $\theta_s$ of probes seen from test zone, the number of selected probes $K$ and the angular coverage of the probe panel seen in test zone. To determine the range $R$, two issues have to be considered. One is the link budget that determines the upper-bound of $R$. The other is the spherical curvature effect that determines the lower-bound of $R$. Readers can refer to \cite{7571141} for detailed discussion on the determination of $R$ for different bands at 2.6, 3.5 and 28\,GHz, where several parameters including the fixed beam power loss are introduced to investigate the spherical curvature effect. It is worth noting that in the \ac{mmWave} frequency bands, the far-field distance gets smaller as wavelength decreases. For example, the Fraunhofer far-field distances \cite{8713575,xuesong_tap} for a 16$\times$16 and an 8$\times$8 \ac{DUT} array (planar and uniformly half-wavelength spaced)  at 28\,GHz are calculated as 0.60 and 0.13\,m, respectively.\footnote{The minimum far-field distance is calculated according to $\frac{2D^2}{\lambda}$, where $D$ is the diameter of the smallest sphere that encloses the radiating elements of the \ac{DUT}. However, the exact antenna size of the \ac{DUT} is usually unknown, and the radiating aperture is also determined by the ground coupling effects. If the \ac{DUT} is viewed as a ``black-box'', the largest device dimension could be utilized. Using this ``black-box'' approach is overkilling and may bring up unnecessary costs on testing hardware, because antennas can only be placed in limited area, e.g., in several possible locations of a device and its configuration complexity is significantly constraint in practice. In this paper, the ``white-box'' approach focusing only on the radiating elements is adopted for simplicity and authenticity. 
}
It is thus reasonable to assume that the curvature effect is insignificant with a proper setting of range length. In addition, it has been concluded in \cite{7571141} that the Fraunhofer distance is not a precondition for $R$ in fading testing. In this regard, we do not consider the metrics for evaluating $R$
in this study, and a practical range length, i.e. 2\,m, is set for the investigations in the sequel. To evaluate the \ac{OTA} emulation performance with different settings of probe number and angular spacing, four evaluation metrics introduced in \cite{8421660} are adopted. For the sake of completeness, definitions of the four metrics are briefly included as follows.

\textit{1) Spatial Correlation Error $e_\rho$:} This parameter $e_\rho$ is to compare the target \ac{PAS} and the emulated \ac{PAS} by comparing the target spatial correlation and the emulated spatial correlation, which is formatted as
\begin{equation}
 \begin{aligned}
e_\rho = \sqrt{ \frac{1}{N^2}\sum_{a=1}^{N} \sum_{b=1}^{N} | \rho_{a,b} -\hat{\rho}_{a,b}|^2  \max{(|\rho_{a,b}|,|\hat{\rho}_{a,b}|)}}.
 \label{eq:spatial_corr_error}
 \end{aligned}
\end{equation} Note that the weight $\max{(|\rho_{a,b}|,|\hat{\rho}_{a,b}|)}$ is applied for emphasizing the deviation of a large correlation coefficient. The reason is that a correlation deviation of a large correlation coefficient has more significant impact than that of a small correlation coefficient \cite{8421660}.

\textit{2) Total Variation Distance of \ac{PAS} $d_p$:} Similar to $e_\rho$, this parameter $d_p$ is introduced as an alternative to compare the target and emulated \ac{PAS}s by comparing the obtained Bartlett beamforming spectra (normalized) of target and emulated channels. Specifically, the Bartlett beamforming spectra $P_t$ and $P_o$ for the target channel and the emulated channel respectively can be formatted as
\begin{equation}
 \begin{aligned}
P_{t/o}(\Omega) = \mathbf a^{H}(\Omega) \mathbf R_{t/o} \mathbf a(\Omega)
 \label{eq:bartlett_pas}
 \end{aligned}
\end{equation} {where $\mathbf R_{t/o} \in \mathbb{C}^{N\times N}$ is the covariance matrix with its $(a,b)$th element equal to the unnormalized $\rho_{a,b}$ in \eqref{eq:target_spatial_corr} for $\mathbf  R_t$ and the unnormalized $\hat{\rho}_{a,b}$ in
\eqref{eq:realized_spatial_corr} for $\mathbf  R_o$,} and $\mathbf a$ is the steering vector.
The total variation distance of \ac{PAS} is then calculated as
\begin{equation}
 \begin{aligned}
d_p = \frac{1}{2}\int \left|   \frac{P_t(\Omega)}{\int P_t(\Omega') d \Omega'}  -   \frac{P_o(\Omega)}{\int P_o(\Omega') d \Omega'}  \right| d \Omega.
 \label{eq:pas_distance}
 \end{aligned}
\end{equation} This can be interpreted as the difference between two 2D joint-distribution functions. The value of $d_p$ ranges between $[0,1]$, with 0 indicating full similarity and 1 full dissimilarity.

\textit{3) Total Variation Distance of Beam Allocation Distributions $d_{\text{ba}}$:}
For the \ac{5G} devices performing beam operations such as beam sweeping and refinement \cite{Giordani2019,3GPPNR}, several beams with indices $\{1, \cdots, B\}$ are pre-defined, and the one with the maximum power is selected. For a stochastic channel with a certain \ac{PAS}, the probability of the $b$th beam is selected should be a certain value. In other words, there is a certain beam allocation distribution for the target channel which indicates each beam's selection probability. The parameter $d_{\text{ba}}$ is exploited to calculate the distance between the beam allocation distributions $p_t$ and $p_o$ of the target and emulated channels as
\begin{equation}
 \begin{aligned}
   d_\text{ba} = \frac{1}{2}\sum_{b=1}^{B} | p_t(\Omega_b) - p_o(\Omega_b) |
 \label{eq:Beam_allocation_distance}
 \end{aligned}
\end{equation} where $\Omega_b$ is the steered direction of the $b$th pre-defined beam. It can be known that the value of $d_{\text{ba}}$ is between $[0,1]$ with $0$ indicating the same distribution.

\textit{4) Beam Peak Distance $d_{\text{bp}}$:} This parameter is similar to $d_{\text{ba}}$, which is used to compare the expected beam directions of $p_t$ and $p_o$ as
\begin{equation}
 \begin{aligned}
   d_\text{bp} = \left|     \sum_{b=1}^{B} \Omega_b p_t(\Omega_b) - \sum_{b=1}^{B} \Omega_b p_o(\Omega_b) \right|.
 \label{eq:Beam_peak_distance}
 \end{aligned}
\end{equation}

In the sequel, investigations on probe number and angular spacing of probes are conducted using the above four parameters as evaluation criteria.

\subsection{Numerical Investigations\label{sec:simulations}}
In this section, simulations are implemented to investigate how different numbers of probes and angular spacing affect the one-cluster channel emulation performance for \ac{DUT}s of different sizes. Specifically, half-wavelength spaced uniform planar arrays at 28\,GHz are considered with fixed array element (isotropic radiation pattern) and array dimensions ranging from 2$\times$2 to 16$\times$16. The probe number $K$ is set to change from 1 to 10, and the angular spacing $\theta_s$ is set to vary from 1$^\circ$ to 20$^\circ$. The range $R$ is fixed to 2\,m as discussed in Sect.\,\ref{sec:metrics}. Moreover, the azimuth and elevation spreads of the cluster are set to 5$^\circ$ and $3^\circ$, respectively, and the cluster \ac{PAS} obeys the Laplace distribution as standardized in \cite{3GPP38901}. Table \ref{tab:simulation_parameters} summarizes the simulation parameters.

\begin{table}[t]
\centering
\caption{\ac{OTA} parameters applied in the simulations}
\scalebox{0.7}{
\begin{tabular}{lclc}
\hline\hline
\multicolumn{4}{c}{{{{\textit{Simulation parameters}}}}}\\\hline
Probe number $K$ & [1:10] & Center frequency& $28$\,GHz     \\
Angular spacing $\theta_s$ & [1:20]  & Cluster azimuth spread & 5$^\circ$  \\
\ac{DUT} array dimension & 2$\times$2 to 16$\times$16 & Cluster elevation spread & 3$^\circ$  \\
\ac{DUT} beam sweeping & \ac{DUT} HPBW  & Cluster \ac{PAS} distribution & Laplace\\
Range length $R$ & $2$\,m & Cluster cases & Aligned/far-nonaligned  \\
\hline\hline
\end{tabular}}
\label{tab:simulation_parameters}
\end{table}

\begin{figure}
  \centering
  \includegraphics[width=0.45\textwidth]{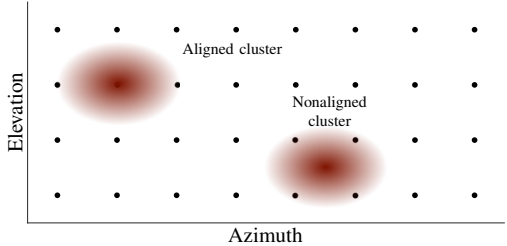}{
  \psfrag{C}[l][l][0.6]{Aligned cluster}
  \psfrag{D}[c][c][0.6]{\tabincell{c}{Nonaligned\\cluster}}
  \psfrag{Azimuth}[c][c][0.8]{Azimuth}
  \psfrag{Elevation}[c][c][0.8]{Elevation}
  }
  \caption{An illustration of aligned and nonaligned cases when cluster evolves in angular domain.}
  \label{fig:cluster_cases}
\end{figure}

In a realistic channel, the dominant cluster usually evolves in the spatial domain. In other words, the channel is dynamic and spatially consistent (e.g., see the realistic indoor channel presented in Sect.\,\ref{secIV:realistic_channel}). Therefore, it is essential that the emulation system can handle the dynamic behaviors with spatial evolution. As illustrated in Fig.\,\ref{fig:cluster_cases}, it can be observed that during the cluster evolution in angular domain, it is possible that the cluster's centroid would align with a probe and also aloof from all probes. We denote them as ``aligned'' case and ``nonaligned'' case, respectively. It is intuitive that the \ac{PAS} may be better generated with less probes in the aligned-case. However, it may be more difficult to mimic the target \ac{PAS} distribution in the non-aligned-case where there is no probe existing at its distribution center, especially in the ``far-nonaligned'' case where the cluster centroid is in the center of a neighbouring four-probes as indicated in Fig.\,\ref{fig:cluster_cases}. Thus, the two very different cases are further considered in the simulation. Another issue needs to be considered in the simulation is how to sweep beams (determine the angular spacing of the beam sweeping), which is related to the calculation of
$d_\text{ba}$ and $d_\text{bp}$.
In the simulation, it is set as the \ac{HPBW} of the \ac{DUT}'s steering beams so that all the \ac{PAS} power can be appropriately covered avoiding too much overlapping. In total 6000 (15$\times$10$\times$20$\times$2) combinations are simulated considering array dimension, probe number, angular spacing and the two cluster-evolution cases. For each combination, the four performance evaluation metrics as elaborated in Sect.\,\ref{sec:metrics} are calculated. Due to the space limitation, representative figures for the results of three different \ac{DUT} dimensions, i.e., 2$\times$2, 8$\times$8 and 16$\times$16, are presented. The $2\times2$ is considered as a typical \ac{UE} device, 8$\times$8 a typical \ac{BS} array, and 16$\times$16 a device with enhanced capability.

\begin{figure}
  \centering
  \includegraphics[width=0.45\textwidth]{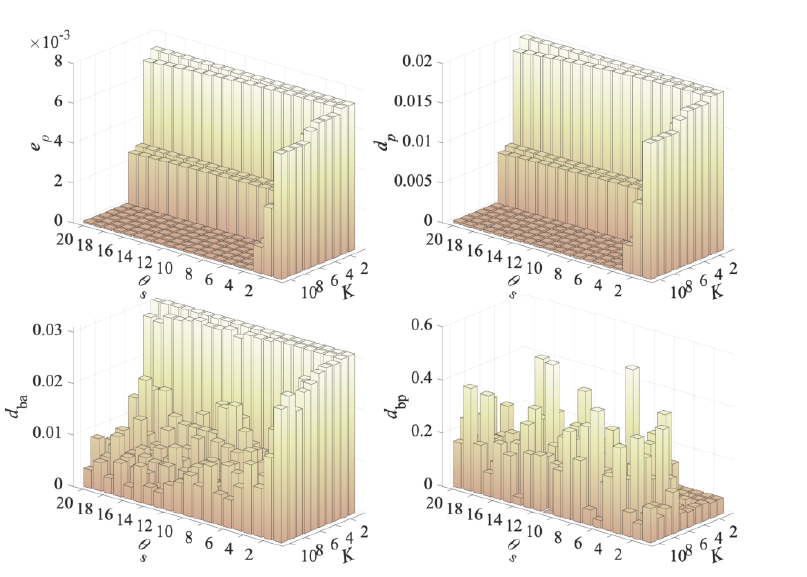}
  \caption{\Ac{SMPAC} \ac{OTA} emulation performance for a 2$\times$2 \ac{DUT} array with different numbers of probes and angular spacing in aligned cluster case.}
  \label{fig:2by2_center}
\end{figure}

\begin{figure}
  \centering
  \includegraphics[width=0.45\textwidth]{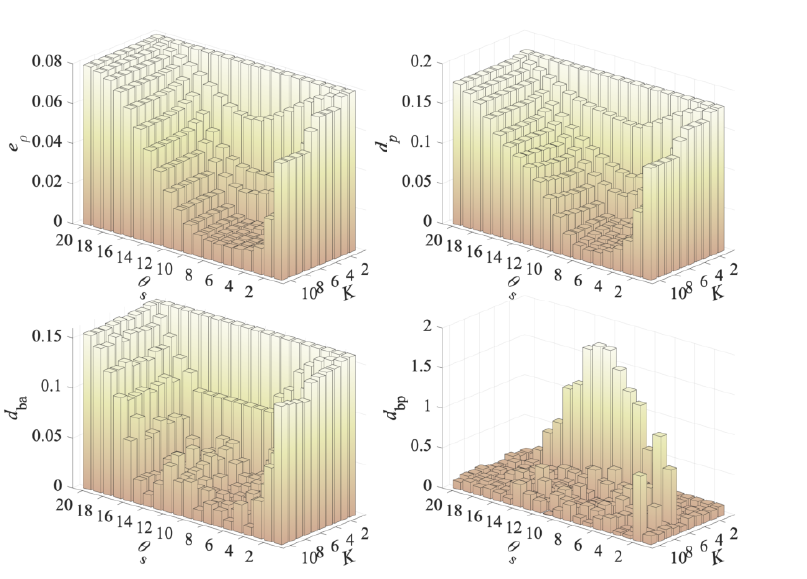}
  \caption{\Ac{SMPAC} \ac{OTA} emulation performance for an 8$\times$8 \ac{DUT} array with different numbers of probes and angular spacing in aligned cluster case.}
  \label{fig:8by8_center}
\end{figure}

\begin{figure}
  \centering
  \includegraphics[width=0.43\textwidth]{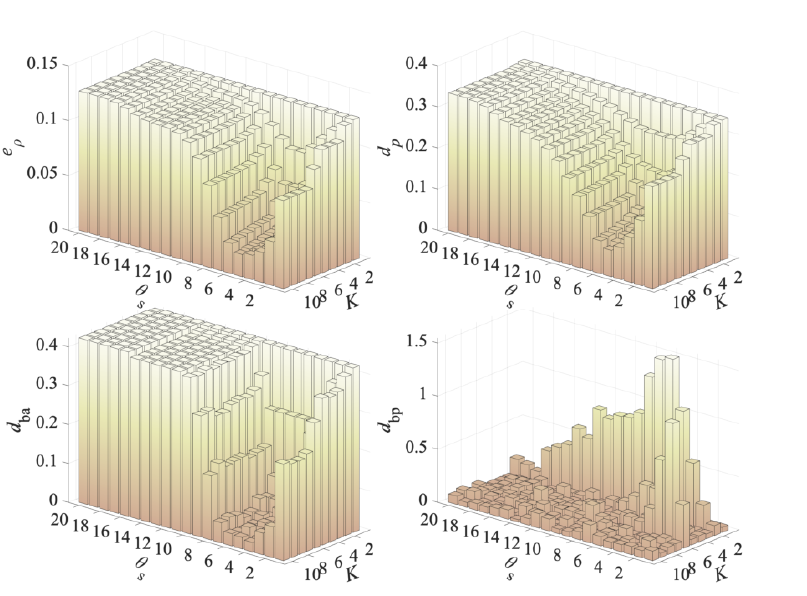}
  \caption{\Ac{SMPAC} \ac{OTA} emulation performance for a 16$\times$16 \ac{DUT} array with different numbers of probes and angular spacing in aligned cluster case.}
  \label{fig:16by16_center}
\end{figure}

\textit{1) Aligned cluster case:} Figs.\,\ref{fig:2by2_center}-\ref{fig:16by16_center} illustrate the four performance evaluation metrics corresponding to different probe number $K$ and angular spacing $\theta_s$ in the case of aligned cluster for a 2$\times$2, 8$\times$8 and 16$\times$16 \ac{DUT} array, respectively. It can be observed from
Fig.\,\ref{fig:2by2_center} that the four performance evaluation metrics are sufficiently small for all $K$-$\theta_s$ pairs. This demonstrates that one probe is adequate to emulate one cluster for a 2$\times$2 \ac{DUT} array. It is reasonable because the beam resolution of this array is limited, and the cluster spread at \ac{mmWave} bands is usually small. With a probe existing at the centroid of the cluster, the \ac{PAS} can be well emulated for the \ac{DUT} array. As illustrated in Fig.\,\ref{fig:8by8_center}, the performance degrades for an 8$\times$8 array, since its ability to resolve different propagation paths is better compared to the 2$\times$2 array. This is similarly true for the $16\times16$ \ac{DUT} array where the performance degrades with the same $K$-$\theta_s$ compared to that of $2\times2$ and
$8\times8$ arrays. Nevertheless, it can be observed from Figs.\,\ref{fig:2by2_center}-\ref{fig:16by16_center} that the performances for the three \ac{DUT}s are still decently well even though only one probe is deployed, as indicated by the maximum values of
$e_\rho$, $d_p$, and $d_\text{ba}$ under $0.4$. It is due to the fact that a probe is aligned to the cluster. Moreover, a common pattern can be observed from Figs.\,\ref{fig:2by2_center}-\ref{fig:16by16_center}. That is, the best performance is obtained with a moderate angular spacing. In other words, when $\theta_s$ is too small or too large, the performance degrades. The reason is that with a too small $\theta_s$, the emulated power is too concentrated; while with a too large $\theta_s$, probes except the aligned one are far away from the main paths of this cluster. Both two cases result in a bad emulation for the \ac{PAS} distribution.

\begin{figure}
  \centering
  \includegraphics[width=0.44\textwidth]{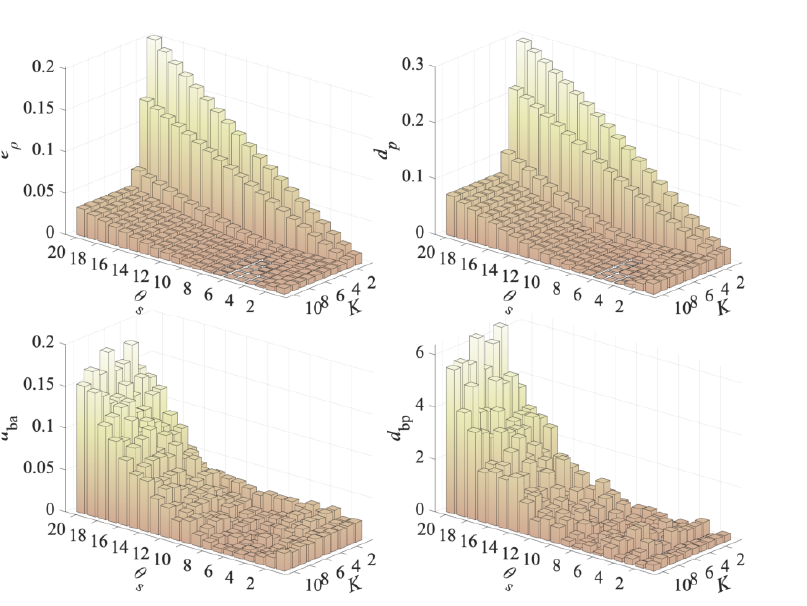}
  \caption{\Ac{SMPAC} \ac{OTA} emulation performance for a 2$\times$2 \ac{DUT} array with different numbers of probes and angular spacings in far-nonaligned cluster case.}
  \label{fig:2by2_noncenter}
\end{figure}

\begin{figure}
  \centering
  \includegraphics[width=0.45\textwidth]{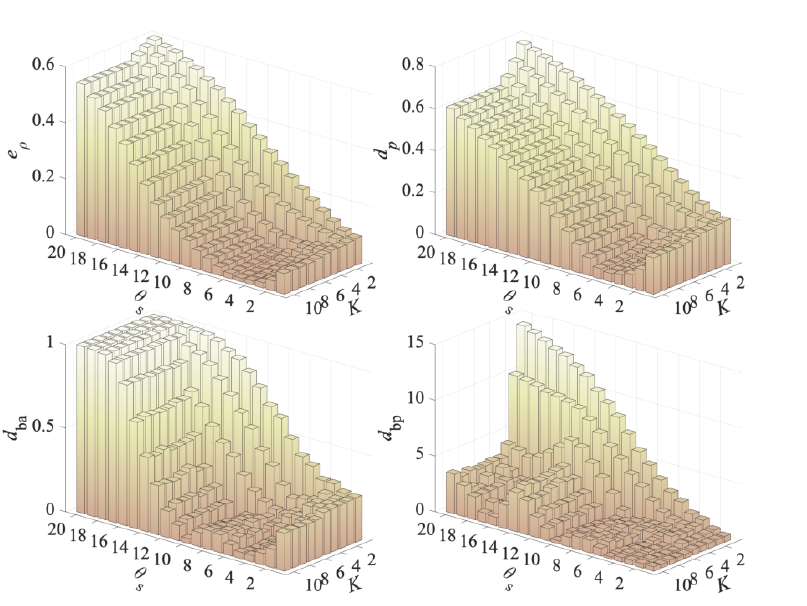}
  \caption{\Ac{SMPAC} \ac{OTA} emulation performance for an 8$\times$8 \ac{DUT} array with different numbers of probes and angular spacings in far-nonaligned cluster case.}
  \label{fig:8by8_noncenter}
\end{figure}

\begin{figure}
  \centering
  \includegraphics[width=0.45\textwidth]{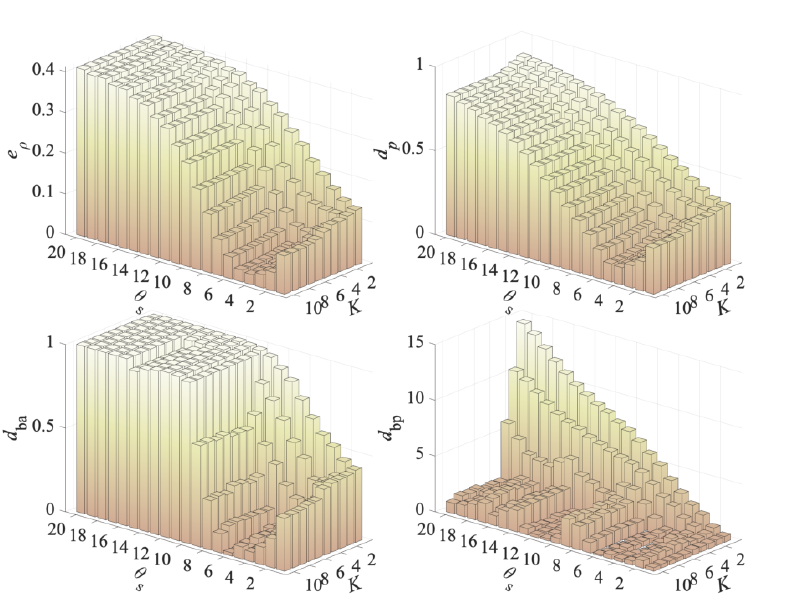}
  \caption{\Ac{SMPAC} \ac{OTA} emulation performance for a 16$\times$16 \ac{DUT} array with different numbers of probes and angular spacings in far-nonaligned cluster case.}
  \label{fig:16by16_noncenter}
\end{figure}

\textit{2) Far-nonaligned cluster case:} Figs.\,\ref{fig:2by2_noncenter}-\ref{fig:16by16_noncenter} illustrate the four performance evaluation metrics corresponding to different probe number and angular spacing in the case of far-nonaligned cluster for a 2$\times$2, 8$\times$8 and 16$\times$16 \ac{DUT} array, respectively. Different from the aligned cluster case, it can be observed from Figs.\,\ref{fig:2by2_noncenter}-\ref{fig:16by16_noncenter} that the errors/distances between the target and the emulated metrics for the non-aligned cluster case are obviously larger (e.g. see the values with one probe where \ac{PAS} ). This is reasonable due to the fact the cluster centroid is far away from any probes, and a better emulation of the \ac{PAS} requires more probes compared to that of the aligned cluster case. Moreover, similar observations as the aligned cluster case can be appreciated from Figs.\,\ref{fig:2by2_noncenter}-\ref{fig:16by16_noncenter} as follows. A larger sized \ac{DUT} requires more probes and smaller angular spacing to gain similar performance to that of a smaller sized \ac{DUT}. With the same angular spacing, a larger number of probes can result in a better performance. However, with the same probe number, a moderate angular spacing is preferable.

\begin{table*}
	\caption{\ac{OTA} setup recommended for different \ac{DUT}s and for different performance levels.}
	\begin{center}
\scalebox{0.75}{
		\begin{tabular}{c|c|c|c|c|c|c|c|c|c|c|c|c|c|c|c}
			\hline  \diagbox{Bounds}{($K$, $\theta_s$[$^\circ$])}{\ac{DUT}}  & 2$\times$2 & 3$\times$3 & 4$\times$4 & 5$\times$5 & 6$\times$6 & 7$\times$7 & 8$\times$8 & 9$\times$9 & 10$\times$10 & 11$\times$11  &  12$\times$12 & 13$\times$13 & 14$\times$14 & 15$\times$15 & 16$\times$16  \\ \hline
      0.10$\times$3    & (1,\,10)   & (1,\,6)   & (1,\,4)   & (1,\,2)   & (2,\,1)   & (3,\,7)   & (3,\,7)   & (4,\,8)   & (4,\,7)   & (4,\,6)   & (4,\,6)   & (4,\,6)   & (4,\,6)   & (4,\,5)   & (5,\,5) \\ \hline
 0.15$\times$3    & (1,\,16)   & (1,\,10)   & (1,\,7)   & (1,\,5)   & (1,\,4)   & (1,\,2)   & (1,\,1)   & (3,\,8)   & (3,\,7)   & (3,\,6)   & (3,\,6)   & (4,\,7)   & (4,\,6)   & (4,\,6)   & (4,\,6) \\ \hline
 0.20$\times$3    & (1,\,20)   & (1,\,14)   & (1,\,10)   & (1,\,7)   & (1,\,6)   & (1,\,4)   & (1,\,4)   & (1,\,1)   & (2,\,5)   & (2,\,5)   & (2,\,5)   & (3,\,5)   & (3,\,5)   & (3,\,5)   & (3,\,5) \\ \hline
 0.25$\times$3    & (1,\,20)   & (1,\,17)   & (1,\,12)   & (1,\,10)   & (1,\,8)   & (1,\,6)   & (1,\,5)   & (1,\,3)   & (1,\,3)   & (1,\,3)   & (1,\,2)   & (3,\,6)   & (3,\,6)   & (3,\,6)   & (3,\,6) \\ \hline
 0.30$\times$3    & (1,\,20)   & (1,\,20)   & (1,\,15)   & (1,\,12)   & (1,\,9)   & (1,\,7)   & (1,\,7)   & (1,\,4)   & (1,\,4)   & (1,\,4)   & (1,\,4)   & (2,\,4)   & (2,\,4)   & (2,\,4)   & (2,\,4) \\ \hline
		\end{tabular}
}
	\end{center}
	\label{tab:guideline}
\end{table*}

To provide guidelines for selecting proper $K$ and $\theta_s$, upper-bounds are defined for the four performance evaluation metrics. With certain predefined bounds, multiple $K$-$\theta_s$ pairs may meet the requirements. Among all the candidates, the pairs with the smallest $K$ are firstly searched, and then the pair with the largest $\theta_s$ is finally chosen as a ``good'' option, since it is reasonable to assume the fading emulator resource is much more expensive, and a larger $\theta_s$ can result in a larger angular coverage. Table\,\ref{tab:guideline} summarizes the recommended (``good'') options for the different \ac{DUT}s and different bounds. Note that the following {three} aspects are considered when generating this table. {\textit{i)} Both aligned and nonaligned conditions are considered.} \textit{ii)} Beam peak distance $d_{\text{bp}}$ is not considered because its value is not confined in a certain range as the other three metrics in [0,1], and it is related to $d_{\text{ba}}$. \textit{iii)} It is possible that one metric slightly exceeds its upper bound, yet other metrics are well under their corresponding bounds. Thus a soft total bound is applied. That is, a setup is considered a candidate if the sum of the metrics' values are smaller than the sum of their bounds. For example, 0.1$\times$3 in Table\,\ref{tab:guideline} indicates the soft bound by summing the three bounds (all set as 0.1) of
$e_\rho$, $d_p$ and $d_\text{ba}$. It can be observed from Table\,\ref{tab:guideline} that with a larger upper bound, basically less probes and/or a larger angular spacing can be applied. For a typical 8$\times$8 \ac{BS} \ac{DUT} array at 0.1$\times$3 soft bound, (3, 7$^\circ$) is recommended, and for a 16$\times$16 \ac{DUT} array (5, 5$^\circ$) is recommended which can be considered as a setup applicable for all \ac{DUT}s.

\textit{3) A cost-effective design principle for dynamic cluster emulation:}

\begin{figure}
  \centering
  \includegraphics[width=0.45\textwidth]{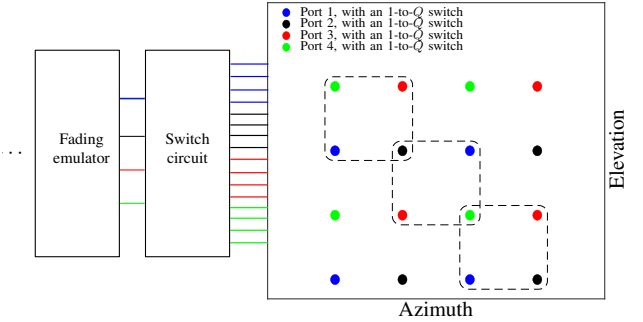}{
  \psfrag{data1}[l][l][0.5]{Port 1, with an 1-to-$Q$ switch}
  \psfrag{data2}[l][l][0.5]{Port 2, with an 1-to-$Q$ switch}
  \psfrag{data3}[l][l][0.5]{Port 3, with an 1-to-$Q$ switch}
  \psfrag{data4}[l][l][0.5]{Port 4, with an 1-to-$Q$ switch}
  \psfrag{Azimuth}[c][c][0.8]{Azimuth}
  \psfrag{Elevation}[c][c][0.8]{Elevation}
  \psfrag{A}[c][c][0.6]{ \begin{tabular}{@{}c@{}} Switch\\circuit\end{tabular} }
  \psfrag{B}[c][c][0.6]{ \begin{tabular}{@{}c@{}} Fading\\emulator\end{tabular} }
  \psfrag{C}[c][c][0.8]{$\cdots$}
  }
  \caption{An sketch of the proposed switch and panel design strategy.}
  \label{fig:panel_design}
\end{figure}

Since the number of active probes required for emulating a cluster is limited, and the active probes are confined in a compact angular area, it is unnecessary to use a switch matrix with full freedom. Alternatively, several 1-to-$Q$ switches each connecting to an output port of the fading emulator can be utilized. 
Fig.\,\ref{fig:panel_design} exemplifies the proposed design strategy, where 4 probes ($K=4$) are active; in other words, the fading emulator provides 4 output ports. An 1-to-$Q$ switch ($Q=4$ in the example) is connected to each output port, and each output port can activate one of the $Q$ probes at one time instant. Totally $KQ$ probes are installed on the panel in an interleaved style as sketched in Fig.\,\ref{fig:panel_design}. With such a design the four output ports can follow a dynamic cluster by activating the necessary probes, e.g. as indicated by the dashed squares in Fig.\,\ref{fig:panel_design}.
The angular coverage can be increased by increasing $Q$. With azimuth and elevation coverage required as
$\theta_\text{A}$ and $\theta_\text{E}$ respectively, $Q$ can be approximately calculated as
\begin{equation}
 \begin{aligned}
  Q =  \left\lceil \frac{\lceil \frac{\theta_\text{A}}{\theta_s} + 1 \rceil  \cdot \lceil \frac{\theta_\text{E}}{\theta_s} + 1 \rceil }{K} \right\rceil  %
  \label{eq:qcalculation}
 \end{aligned}
\end{equation} with $\lceil x \rceil$ indicating the smallest integer larger or equal $x$. The advantage of this strategy compared to \cite{8421660} is that a switch circuit with full freedom is replaced with several 1-to-$Q$ sub-switches (implemented e.g. using multistage 1-to-$Q^\prime$ switches), thus significantly decreasing the complexity and cost.

\begin{figure}
  \centering
  \subfigure[]{\includegraphics[width=0.24\textwidth]{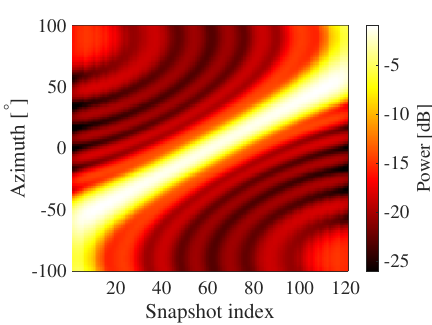}}
  \subfigure[]{\includegraphics[width=0.24\textwidth]{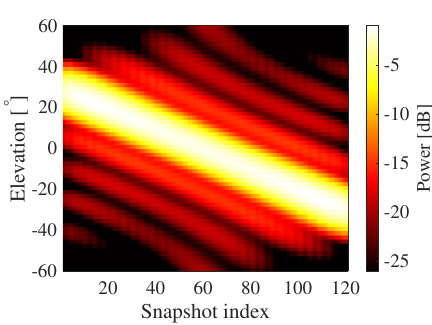}}
  \caption{Target dynamic \ac{PAS}s for the artificial dynamic channel. (a) Dynamic azimuth \ac{PAS}. (b) Dynamic elevation \ac{PAS}.}
  \label{fig:target_dynamic_spectrum}
\end{figure}

\begin{figure}
  \centering
  \subfigure[]{\includegraphics[width=0.24\textwidth]{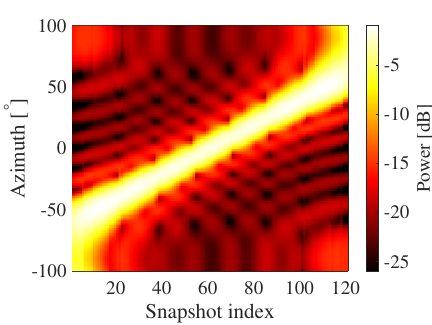}}
  \subfigure[]{\includegraphics[width=0.24\textwidth]{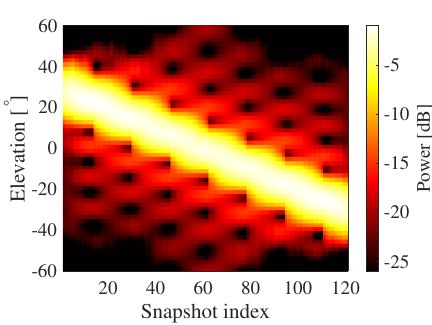}}
  \caption{Emulated \ac{PAS}s for the artificial dynamic channel. (a) Dynamic azimuth \ac{PAS}. (b) Dynamic elevation \ac{PAS}.}
    \label{fig:ota_dynamic_spectrum}
\end{figure}

To evaluate this setup, an artificial dynamic cluster is emulated for an 8$\times$8 \ac{DUT} array with $K=4$ and $\theta_s=8^\circ$. The dynamic cluster centroid is set to evolve simultaneously and
linearly from -60$^\circ$\,to\,60$^\circ$ in azimuth and from -30$^\circ$\,to\,30$^\circ$ in elevation within a few time snapshots. Figs.\,\ref{fig:target_dynamic_spectrum}(a) and (b) illustrate the target \ac{PAS}s sliced in azimuth and elevation planes, respectively, while Figs.\,\ref{fig:ota_dynamic_spectrum}(a) and (b) the emulated ones.\footnote{Note that the target or emulated \ac{PAS} is a 3D spectrum in both azimuth and elevation domains at each time snapshot. It is not easy to plot/show the 3D spectrum dynamically evolving with respect to time. Alternatively, two slices of the 3D spectrum in azimuth and elevation planes respectively (like the E-plane and H-plane of an antenna pattern) are obtained at each time snapshot. By concatenating these slices with respect to the snapshot-index (time), the dynamic \acp{PAS} are shown in Figs.\,\ref{fig:target_dynamic_spectrum} and \ref{fig:ota_dynamic_spectrum}.} Through intuitive visual inspection, the target and emulated dynamic \ac{PAS}s are quite similar. The quantitative similarity for azimuth plane is calculated as 96.7\%
($d_p$=3.3\%) using \eqref{eq:pas_distance}, and that for elevation \ac{PAS}s is calculated as 94.5\%. This demonstrates that the dynamic channel has been emulated with decently good performance using the proposed switching strategy. {It is worth noting that when the cluster is evolving between an aligned cluster and an non-aligned cluster using the same resource, the emulation spectra as illustrated in Fig.\,\ref{fig:ota_dynamic_spectrum} are not very smooth since nonaligned-cluster snapshots are more resource-demanding.}

%
%
%
%
\section{Emulation Validation Using A Realistic Indoor dynamic mmWave Channel\label{secIV:realistic_channel}}
In this section, an indoor measurement campaign is introduced. By exploiting a high-resolution channel parameter estimation algorithm as well as a clustering identification and tracking algorithm, dynamic clusters of the channel are extracted from the measurement data \cite{caidynamic}. The dynamic evolution behavior of the dominant cluster are shown. Furthermore, the applicability of the proposed \ac{OTA} setup is validated by the measure channel. This measurement campaign also sheds lights on how large the probe panel should be designed.

\subsection{Measurement campaign and cluster tracking}

\begin{figure}
\centering
  \includegraphics[width=0.45\textwidth]{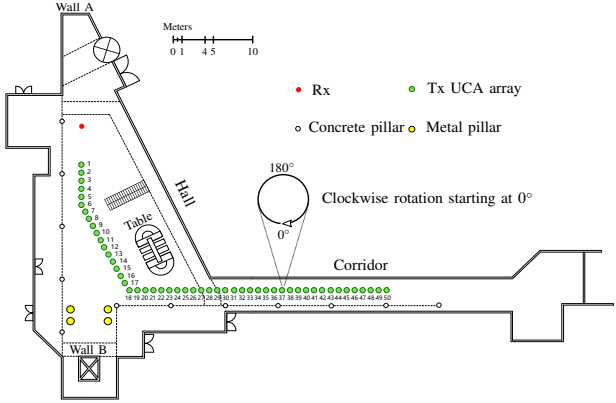}
  { \psfrag{Hall}[c][c][0.6]{Hall}
    \psfrag{g}[c][c][0.6]{Corridor}
    \psfrag{Tx}[c][c][0.6]{Tx}
    \psfrag{Rx}[c][c][0.6]{Rx}
    \psfrag{a}[c][c][0.4]{0}
    \psfrag{b}[c][c][0.4]{1}
    \psfrag{c}[c][c][0.4]{4}
    \psfrag{d}[c][c][0.4]{5}
    \psfrag{e}[c][c][0.4]{10}
    \psfrag{r}[l][l][0.6]{Rx}
    \psfrag{t}[l][l][0.6]{Tx \ac{UCA} array}
    \psfrag{w}[l][l][0.6]{Concrete pillar}
    \psfrag{y}[l][l][0.6]{Metal pillar}
    \psfrag{n}[l][l][0.6]{Clockwise rotation starting at 0$^\circ$}
    \psfrag{o}[c][c][0.5]{0$^\circ$}
    \psfrag{l}[c][c][0.5]{180$^\circ$}
    \psfrag{M}[c][c][0.4]{Meters}
    \psfrag{L}[c][c][0.5]{Table}
    \psfrag{i}[c][c][0.5]{Wall A}
    \psfrag{J}[c][c][0.5]{Wall B}
    }
    \caption{The layout of the indoor measurement scenario.}
    \label{fig:indoor_sce}
\end{figure}

An ultra-wideband measurement system developed based on a \ac{VNA} and the radio-over-fiber technique was applied in the measurement campaign. Phase coherent measurements can be done by exploiting the proposed phase compensation scheme using optical circulators. Furthermore, a dynamic range of 112\,dB at 30\,GHz can be obtained with an optical fiber of 300\,m length due to its low attenuation to the signals. Readers are referred to \cite{8901446} for the system details. The measurement campaign was conducted in an indoor hall-to-corridor scenario with its layout illustrated in Fig.\,\ref{fig:indoor_sce} (Fig.\,1(a) in \cite{caidynamic}), and the ceiling height was about 10\,m. Tables, stairs, metallic pillars, concrete pillars, etc. existed in the hall. Two quasi-omnidirectional bi-conical antennas were utlized as Tx and Rx antennas, respectively. The Rx antenna was fixed during measurement as indicated in Fig.\,\ref{fig:indoor_sce}, with a height of 3\,m. The Tx antenna was installed on a rotator with a height of 1.15\,m, and it was rotated in the azimuth plane with 360 uniform steps to perform a virtual \ac{UCA} measurement where the circular radius was 0.25\,m. In each step, \ac{CTF} between Tx and Rx was measured using the \ac{VNA} sweeping 2000 samples in the 28-30\,GHz frequency band. Totally 50 \ac{UCA} locations were measured form the hall to the corridor.

Based on the measured \ac{CTF} at each \ac{UCA} location, the propagation parameters of \acp{MPC}, including propagation delays, azimuth angles, elevation angles and complex amplitudes, are estimated using a high-resolution estimation algorithm \cite{8713575,xuesong_tap}. Note that since the array measurement was applied only for one side, the angle information at the other side cannot be obtained. The underlying signal model of the channel impulse response can be formatted as
\begin{equation}
\begin{aligned}
h(\tau, \phi, \theta, d ) = \sum_{\ell=1}^{L} \alpha_\ell \delta(\tau-\tau_\ell)     \delta(\phi-\phi_\ell) \delta(\theta-\theta_\ell)  \delta(d-d_\ell)
\end{aligned}
\end{equation} where $L$ is the total number of \acp{MPC}, and $\alpha_\ell$, $\tau_\ell$, $\phi_\ell$, $\theta_\ell$ and $d_\ell$ represent the complex amplitude, propagation delay, azimuth, elevation and spherical wavefront distance of the $\ell$th \ac{MPC}, respectively. Furthermore, based on the \ac{MPC} estimation results, clusters of \acp{MPC} are grouped at each location by exploiting a threshold-based clustering algorithm, and dynamic clusters are associated across the 50 locations using a cluster tracking algorithm as proposed in \cite{caidynamic}. For the \ac{MPC} estimation, cluster identification and cluster tracking algorithms, readers are referred to
\cite{8713575,xuesong_tap,caidynamic} for more details.

%

\subsection{\ac{OTA} Emulation for the dynamic channel}

\begin{figure}
\centering
  \includegraphics[width=0.45\textwidth]{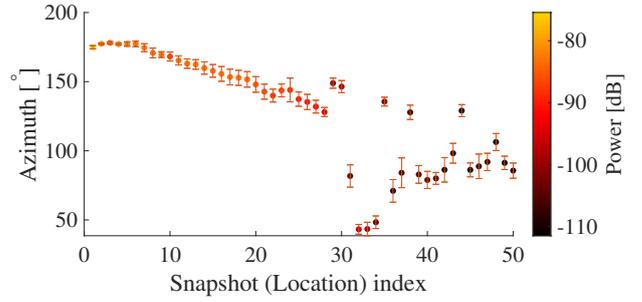}
  {\psfrag{Snapshot index}[c][c][0.8]{Snapshot (Location) index}}
    \caption{The dynamic evolution of the dominant cluster for the indoor channel.}
    \label{fig:dynamic_cluster_indoor}
\end{figure}

\begin{figure}
  \centering
  \subfigure[]{\includegraphics[width=0.24\textwidth]{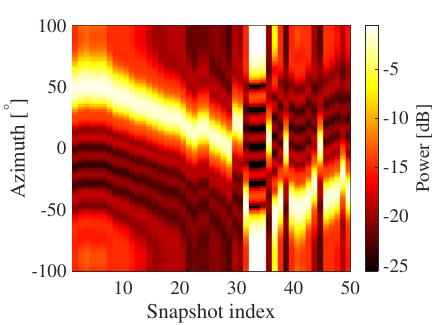}}
  \subfigure[]{\includegraphics[width=0.24\textwidth]{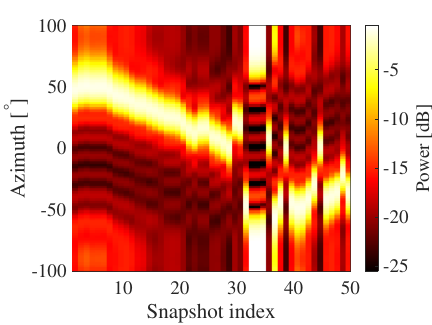}}
  \caption{Target and emulated dynamic \ac{PAS} for the measured indoor channel. (a) Target azimuth \ac{PAS}. (b) Emulated azimuth \ac{PAS}.}
  \label{fig:indoor_pas_comparison}
\end{figure}

Fig.\,\ref{fig:dynamic_cluster_indoor} illustrates the dynamic evolution of the dominant cluster in azimuth domain obtained from the cluster identification and tracking results across the 50 locations, and the error bars indicate the azimuth spread at each location. The mean and standard deviation of the azimuth spread are calculated as 4.7$^\circ$ and 2.2$^\circ$, respectively. It can be observed that in the hall scenario (locations 1-28), the dominant cluster is rather stable with relatively high power. It is actually the \ac{LoS} cluster. After entering into the corridor, the \ac{LoS} cluster was blocked, and the dominant cluster changed abruptly to another \ac{NLoS} ones with very different azimuths. Furthermore, the cluster was with much weaker power, and its evolution was less stable than that of \ac{LoS} cluster. This demonstrates that beam tracking is essential for the device to track promptly the change of the dominant cluster. Meanwhile, beamforming is critical to mitigate the attenuation to achieve acceptable link budget at \ac{mmWave} frequency bands especially in the \ac{NLoS} case (corridor). All these put significant importance on the channel emulation of the realistic channels for performance evaluation. Figs.\,\ref{fig:indoor_pas_comparison}(a) and (b) illustrate the target indoor azimuth \ac{PAS} and emulated azimuth \ac{PAS} by exploiting the setup with $K=4$ and $\theta_s=8^\circ$ for an 8$\times$8 \ac{DUT} array. Note that the cluster power has been normalized at each location. It can be observed that the two \ac{PAS}s are similar, and the similarity defined as $1-d_p$ [see \eqref{eq:pas_distance}] is calculate to be 98.2\%.
It is worth noting that the azimuth coverage of the panel should be at least around 150$^\circ$ since the dominant cluster changed in a azimuth range of around 150$^\circ$ as illustrated in Fig.\,\ref{fig:dynamic_cluster_indoor}. Moreover, it can be observed that the cluster azimuth range was not centered at 0$^\circ$, thus a pre-rotation should be applied to the device to make the center of the dominant cluster's angle range align with the probe panel center. 


%

\section{Conclusions\label{section:conclusion}}

In this paper, a dynamic \ac{mmWave} channel emulation method for \ac{5G} \ac{mmWave} massive \ac{MIMO} devices in a sectored multi-probe anechoic chamber (SMPAC) setup was investigated. {We exploited the fact that with beamforming, only a few dominant clusters or even one dominant cluster have to be considered in the \ac{OTA} emulation.}  
Based on this dominant-clusters concept, a cost-efficient \ac{OTA} emulation strategy for dynamic \ac{mmWave} channels was proposed. That is, we focused on reproducing the dominant cluster(s) with high accuracy rather than the composite channel. A switch-circuit with low cost profile using interleaved probes that belong to different 1-to-multiple sub-switches was also designed for this purpose. Comprehensive simulations demonstrated that 4 probes are adequate to simulate the dynamic \ac{mmWave} channels with high accuracy (errors below 0.1) considering the dominant-cluster properties, although the optimized angular spacing among probes needs to be set with respect to particular \ac{DUT} size (see Tabel\,\ref{tab:guideline}). Moreover, the dynamic \ac{mmWave} channel measured in an indoor scenario showed that the dominant cluster evolved in a relatively large azimuth range (i.e. around 150$^\circ$) and presented abrupt changes, and its mean azimuth spread was observed to be around 4.5$^\circ$. The proposed \ac{SMPAC} setup was able to emulate the dynamic behaviors of the measured reality with high accuracy, which further validated the proposed strategy. This work can serve as a certain guideline for \ac{OTA} testing of \ac{5G} devices operating under \ac{mmWave} massive \ac{MIMO} channel conditions.

\setlength{\itemsep}{-20em}
\renewcommand{\baselinestretch}{0.8}

\bibliographystyle{IEEEtran}
\bibliography{ota_reference}

\end{document}